# Comet 21P/Giacobini-Zinner: Narrowband Photometry of the Prototype of Carbon-Chain Depleted Comets at Multiple Apparitions

David G. Schleicher

Lowell Observatory, 1400 W. Mars Hill Rd, Flagstaff, Arizona 86001, USA

## ABSTRACT

We obtained extensive narrowband photoelectric photometry of Comet 21P/Giacobini-Zinner with observations spanning 33 years. The original data from 1985 (Schleicher et al. 1987) were re-reduced and are presented along with data from three additional apparitions including 2018/19. The original conclusion regarding Giacobini-Zinner's chemical composition remains unchanged, with it having a 4-6× depletion in the carbon-chain molecules $C_2$ and $C_3$, and in NH, as compared with both OH and CN. The comet continues to exhibit a large asymmetry in production rates as a function of time and heliocentric distance, with production reaching a peak 3-5 weeks prior to perihelion. All species, including dust, follow the same general production rate curve each apparition, and the carbon-bearing species are always very similar to one another. However, OH and NH each differ in detail from the carbon-bearing species, implying somewhat varied composition between source regions. Longer term, there are only small secular changes among the apparitions before and near perihelion, but larger changes are evident as the comet recedes from the Sun, suggestive of a progressive precession of the rotation axis.

Subject headings: comets: general – comets: individual (21P/Giacobini-Zinner) – methods: data analysis – methods: observational

Unified Astronomy Thesaurus concepts: Comets (280); Short period comets (1452); Comet nuclei (2160); Comae (271); Comet volatiles (2162); Near-Earth objects (1092)

## 1. INTRODUCTION

Discovered at the start of the last century, Comet 21P/Giacobini-Zinner (hereafter, G-Z) was first identified as having a quite high CN-to-$C_2$ brightness ratio, suggestive of an unusual chemical composition, in a 1913 spectrum as reported by Bobrovnikoff (1927) in his summary of spectra for 22 comets. Nearly 50 years later, Mianes et al. (1960) obtained filter photometry showing that the CN-to-$C_2$ ratio was much higher in G-Z than in Comet 2P/Encke (see also Cochran & Barker 1987), though whether CN was enhanced or $C_2$ was depleted was unknown. By the 1980s, emissions of these and additional molecular species were beginning to be systematically studied in a number of comets by a few teams of researchers. Moreover, based on production rate computations, it was beginning to be quantitatively determined that most comets had fairly similar gas compositions, though the details were uncertain due to the use of differing techniques, model parameters, and which specific objects had been measured by each group. Regardless, G-Z's unusual chemical composition was obvious during its excellent 1985 apparition, with Cochran & Barker (1987) reporting that both $C_2$ and $C_3$ were depleted with respect to CN by about a factor of five. Somewhat larger depletions were also determined spectroscopically by Beaver et al. (1990), who additionally determined that $NH_2$ was depleted by about a factor of three. Our own narrowband filter measurements with a photoelectric photometer yielded quite similar results for $C_2$ and $C_3$, but also included NH, whose depletion matched that of $NH_2$, and OH whose abundance ratio to CN was quite normal (Schleicher et al. 1987). Since OH is the direct byproduct of water – the dominant volatile species in most comets – we established that these other species were indeed depleted rather than CN being over-abundant in G-Z.

Our extensive observations in 1985 also revealed an unusual behavior of the gas and dust production rates. In particular, rates peaked approximately four-to-six weeks prior to perihelion, then decreased throughout the perihelion passage to only about one-third at a comparable time and distance outbound, and then remained essentially level to the end of observations five weeks later. Though more limited in temporal extent, this behavior of an early peak in production was confirmed based on OH spectra obtained with the International Ultraviolet Explorer (IUE) satellite by McFadden et al. (1987), who also noted a lack of short-term variability associated with nucleus rotation.

The next practical apparition was in 1998/99, though it favored the pre-perihelion time frame. By this time, A'Hearn et al. (1995) had shown that Giacobini-Zinner was only the first of several comets that exhibited varying degrees of $C_2$ (and usually $C_3$) depletion, and that most of these were in the Jupiter-family dynamical group. As the prototype of an entire compositional class, G-Z thus attracted the attention of researchers using a new generation of instrumentation, particularly in the IR (eg. Weaver et al. 1999; Mumma et al. 2000) in an attempt to measure several parent species. Space-based IUE observations were replaced by SOHO/SWAN measurements of the Lyman-alpha coma (Combi et al. 2011), and narrowband imaging was also obtained at this time (eg. Lara et al. 2003).

The recent 2018/19 apparition brought nearly identical circumstances to that in 1985, though G-Z was upstaged by the ideal passage of 46P/Wirtanen only three months later. A new generation of instrumentation was again brought to bear on G-Z, in particular greatly improved sensitivities in the IR and millimeter regimes (cf. Faggi et al. 2019; Roth et al. 2020; Cordiner 2019), along with high spectral resolution spectroscopy in the visible to measure detailed line intensity ratios (Cochran et al. 2020) and isotope ratios (Moulane et al. 2020). Moulane et al. also obtained extensive narrowband imaging production rate computations, while our own imaging campaign was primarily intended for morphological studies. Specifically, we approached the 2018/19 apparition with two primary goals, one to measure the expected jet morphology associated with isolated source regions that might explain the odd characteristics of G-Z's production rates along its orbit that we had measured in 1985, and the other to greatly extend the heliocentric distance coverage to better constrain the activity on the nucleus as a function of time. Secondarily, we also wanted to search for any evidence of long-term secular changes over a third of a century. Analyses of the imaging data and associated jet modeling are on-going and will be presented separately.

In this paper, we concentrate on production rates along with the chemical composition as a function of heliocentric distance and time obtained using our classical narrowband photometry technique for Giacobini-Zinner. Because photometric reduction procedures with the, at that time, new IHW filters were not yet finalized, nor were model parameters, our 1985 data (Schleicher et al. 1987) have been entirely reprocessed and are presented here in addition to our observations from the more recent apparitions. Also note that, due to our emphasis on imaging at the recent apparition, each observing run consisted of a single night of classical photometry, followed by as many nights as possible of imaging. This resulted in having only one night of photometry per run, unlike prior apparitions, though we indeed greatly extended our coverage of the heliocentric distance range as compared to 1985. Other details of the observations and reductions are presented in Section 2. Results associated with gas and dust production rates as functions of heliocentric distance and of time are presented in Section 3, including examination of any evidence of secular variations over the third of a century our data encompasses, and G-Z's composition is compared to our entire photometric database. Discussion of our findings is presented in Section 4.

## 2. OBSERVATIONS AND REDUCTIONS

### 2.1 Instrumentation

All observations of Comet 21P/Giacobini-Zinner presented here were obtained with conventional photoelectric photometers and narrowband comet filters. The 72-in. (1.8-m) Perkins, the 42-in. (1.1-m) John S. Hall, and the 31-in (0.8-m) telescopes at Lowell Observatory were used in the 1985 and 1998/99 apparitions, while the 24-in (0.6-m) telescope at Perth Observatory was also employed during 1998/99. A single observation obtained in 2012 and all observations in 2018/19 were obtained at the Hall 42-in telescope using the same photometric system as prior apparitions. EMI 6256 photomultiplier tubes (one at each observatory) were used as detectors, in conjunction with a pulse counting system at Lowell and a DC amplifier at Perth. In 1985, a set of the International Halley Watch (IHW) filters was employed (A'Hearn 1991) while all later measurements were made with sets of Hale-Bopp (HB) filters (Farnham et al. 2000). While the same emission bands are isolated by the different generations of filter sets, the wavelengths of the continuum band passes have changed. Specifically, the green continuum centered at 4845 Å for the IHW set was moved to 5260 Å in the HB set, the UV continuum has gone from 3650 to 3448 Å, and a new intermediate continuum filter (blue) at 4450 Å was introduced with the HB set.

### 2.2 Observations

Between one and seven photometric sets were successfully obtained of G-Z on each of 48 nights, resulting in a total of 125 sets over four apparitions, though only one set was attempted in 2012, an apparition with quite poor circumstances. Each set typically consisted of five gas filters, isolating emission from OH, NH, CN, $C_3$, and $C_2$, along with 2–3 continuum filters, but occasionally fewer filters were used due to a lack of time or to the faintness of the comet. The comet was usually observed in each filter for several 10–30 s integrations, and these were either followed by or interspersed with separate sky measurements more than 15 arcmin away from the comet. Observational circumstances for each night are summarized in Table 1, including the heliocentric ($r_H$) and geocentric ($\Delta$) distances as well as the phase angle ($\theta$) of G-Z. The diameter of the photometer entrance aperture ranged between 14 and 205 arcsec, while the corresponding projected aperture radius ($\rho$) varied between 2800 and 148,000 km, though the great majority of observations were obtained with $\rho$ between 5000 and 50,000 km, with a median value of 18,000 km. Note that the apertures employed in 1985 were systematically smaller than in later apparitions due to the f-ratios of the telescopes in use; this is discussed in more detail in Section 3.2. The mid-time and the aperture size for each observational set are listed in Table 2. Comet flux standard stars were measured to determine nightly extinction coefficients and instrumental corrections for each filter.

[TABLE 1 HERE: OBSERVING CIRCUMSTANCES AND L/N]

*2.3. Reductions*

We followed our standard methods for data reduction, as detailed in A'Hearn et al. (1995). For the three apparitions employing the HB filters, data were reduced to fluxes following the specific procedures and using the coefficients detailed in Farnham et al. (2000). The original 1985 data were re-reduced following the revised procedures for the IHW filters detailed in the appendix of Farnham & Schleicher (2005). These revised reductions yield improved decontamination of the green and UV continuum filters by the $C_2$ and $C_3$ emission band wings, respectively, as compared to the original reductions. These changes result in 1985 band fluxes, on average, lower only by about 7% for CN and higher by about 10% for $C_2$; however, the changes are much greater for $C_3$ whose values increase by ~2.1 times due to much more extensive $C_3$ band wings being accounted for (cf. Schleicher & Osip 2002; Schleicher et al. 2003; Farnham & Schleicher 2005). The final band fluxes for OH and NH also changed significantly, but in a less systematic manner than the carbon-bearing species. Contamination of the original UV continuum filter altered the amount of underlying continuum computed for the NH filter, while improved extinction corrections for ozone at the OH band (Farnham & Schleicher 2005) yielded changes as a function of air mass. The resulting emission band flux values for the gas species following continuum subtraction, along with the continuum fluxes, are given in Table 2.

[TABLE 2 HERE: FLUXES & COLUMN ABUNDANCES]

Gas species' fluxes were converted to column abundances by applying the appropriate fluorescence efficiencies (*L/N*). The values for $C_2$ and $C_3$ are given in A'Hearn et al. (1995), while the OH values – that vary with heliocentric velocity ($r_H$) due to the Swings effect – are from Schleicher & A'Hearn (1988). In the cases of CN and NH, the situation is further complicated because, in addition to the basic Swings effect, the fluorescence efficiencies also vary with heliocentric distance due to changing numbers of populated energy levels. Beginning with our analyses of 73P/Schwassmann-Wachmann 3 (Schleicher & Bair 2011), we incorporated the fluorescence computations from Schleicher (2010) for CN and from Meier et al. (1998) for NH. The resulting nightly $L/N$ values for OH, CN, and NH are listed in Table 1.

A standard Haser model (Haser 1957) was used to extrapolate total coma abundances from the column abundances, and gas production rates were computed by dividing by the assumed daughter lifetimes. The parent and daughter Haser scalelengths, as well as daughter lifetimes, are from A'Hearn et al. (1995), with all values assumed to scale as $r_H^2$. These scalelengths also varied from those used in the original 1985 analyses and, coupled with the changes in emission band fluxes previously noted, the results presented here supersede those in Schleicher et al. (1987). The final column abundances ($M(\rho)$), and gas production rates ($Q$) are presented in Tables 2 and 3, respectively. For OH, with a single definitive parent, $H_2O$, we also derived the water production rate using the empirical relation derived by Cochran & Schleicher (1993) and discussed in Schleicher et al. (1998); the values are listed in the last column of Table 3. As a proxy for dust production, we continue to use the quantity $A(\theta)f\rho$, first introduced by A'Hearn et al. (1984), for each continuum filter. The value of this product of the dust albedo at the given phase angle, the filling factor for the aperture, and the projected aperture radius, makes no assumptions regarding the particle size distribution and is independent of wavelength and aperture size if the dust grains are gray in color and have a canonical $1/\rho$ radial distribution. The resulting values of $A(\theta)f\rho$ are also included in Table 3. Finally, because the phase angle varied over a range of 21° to 78°, we also compute the phase adjusted $A(0°)f\rho$ for certain purposes such as the dust-to-gas ratio. While this adjustment has not been applied to the Table 3 values, the adjustment factors from the Schleicher-Marcus composite dust phase curve (see Schleicher & Bair 2011) are given in Table 1.

[TABLE 3 HERE: PRODUCTION RATES]

The uncertainties associated with each data point in Table 3 are 1-σ values derived from the propagation of observational errors based on photon statistics. Since the positive and negative uncertainties in log space are unbalanced but can be readily calculated, we give only the "+" values in the table. In several cases, due either to the faintness of the comet and/or the depletion in some species, the derived flux values were negative following sky removal for the continuum points, or following sky and continuum removal for the emission bands (caused by the uncertainties of each quantity). Since negative flux values are not possible, we set these values to zero for any resulting computations, and the associated logarithms are listed as "undefined" in Tables 2 and 3.

## 3. PHOTOMETRIC RESULTS

### 3.1 Gas Production Rates vs Distance, Time, and Apparition

We begin by examining the overall behavior of Giacobini-Zinner's gas and dust production rates, by plotting in Figure 1 the log of each value as a function of the time from perihelion. Measurements from each apparition are distinguished by different symbols and colors. It is immediately evident that all species exhibit higher production rates prior to perihelion, and that rates peaked about 3-5 weeks before perihelion at each of the three apparitions for which we have sufficient data (see Table 3). While the $A(\theta)f\rho$ results are consistent with the gas in this regard, we defer other discussion of the dust until Section 3.4.

[FIGURE 1 HERE: PRODUCTION RATES VS TIME; 6 PANEL]

We plot the production rates versus time rather than distance because the shape of the heliocentric distance dependence in log-log space is very non-linear. In addition to production rates peaking prior to closest approach, $Q$s drop rapidly during the 4-6 weeks surrounding perihelion (by 3-4× while the heliocentric distance is nearly unchanged) and then nearly flatten out for another 1-2 months before continuing to drop off at larger distances. While this behavior makes the computation of $r_H$-dependencies nearly useless as the values would differ at almost any segment one might choose, a single computation was interesting for bulk comparison purposes between the species. We've simply determined the slope in log-log space for the entire data set, except for the two nights in 1999 and 2019 beyond 2 AU when the comet was too faint for most filters to be used, and excluding highly uncertain data points. As usual for most comets, the three carbon-bearing species in G-Z all have similar slopes, ranging between -2.3 ± 0.4 for $C_3$, -2.6 ± 0.4 for CN, and -3.0 ± 0.3 for $C_2$, and these differences are largely associated with data points at large $r_H$ with high uncertainties. OH and NH, however, were both considerably steeper, with slopes of -4.2 ± 0.4 and -4.5 ± 0.5 (or more, depending on the specific large uncertainty cut-off), respectively. Note that the steeper slopes for OH and NH are also not unusual; for instance, similar behavior was observed in Comets 10P/Tempel 2 (Knight et al. 2012) and 103P/Hartley 2 (Knight & Schleicher 2013), among others. Again, other than showing that the three carbon-bearing species are very similar and somewhat shallower than OH and NH, the *specific values of these slopes are not meaningful* because the actual shape of the curves is very non-linear.

The overall production rate variations with distance and time also yield differing amounts of pre-/post-perihelion asymmetries as a function of distance. For instance, the largest asymmetries occur about 10 weeks from perihelion, when $Q$(CN) on 2018 July 3 ($r_H$ = 1.37 AU) was 8× greater than at the same distance outbound on November 15, while in 1998/99 at a slightly larger distance of ~1.44 AU the asymmetry was 13×. The ratio was smaller but still very significant closer to perihelion; on 2018 August 19 (1.06 AU), $Q$(CN) was 3.1× that on October 6 (1.08 AU). Somewhat less certain, the ratios at extreme distances appear to decrease from the maximum values, with the pre-perihelion $r_H$-dependence steeper than the $r_H$ slope post-perihelion. While the signal-to-noise ratios are significantly poorer for $C_2$ and $C_3$ due to their depletions, the characteristics shown by CN appear to be closely followed. In the case of OH, the general behavior is quite similar to CN, but with a somewhat smaller pre-/post-perihelion ratio near 1.4 AU, while NH has too much scatter to accurately quantify the asymmetry.

With a quite stable orbit for the last half-century, and only a 0.02 AU change in perihelion distance in 2018 from the other apparitions for which we have data (from 1.03 AU to 1.01 AU), G-Z is unremarkable in that there is essentially no secular change in production rates either prior to or near perihelion. Given this result, we were surprised to see significant changes outbound beyond about 30 days (>1.1 AU). While CN, having the best S/N, shows this most clearly and is discussed next in detail, the same basic behavior is evident in all of the gas species and even the dust. Specifically, after the extremely rapid decrease in $Q$ in the ~6 weeks surrounding perihelion, $Q$s were essentially constant from 1.16 AU to 1.45 AU despite the progressive decrease in solar flux. These nearly level or even slightly increasing production rates were originally noted by us (Schleicher et al. 1987), and even though the absolute values of $Q$(CN) in 1985 changed due to the use of differing model parameters (see Section 2.3), this trend with time and distance remains the same. In the 1998/99 apparition, however, while $Q$(CN) initially follows this same near-constant trend to 1.32 AU, $Q$(CN) then abruptly drops by 6× by 1.45 AU, only 2 weeks later, corresponding to the still-level $Q$s of 1985. By the 2018/19 apparition, a nearly 4× drop has already taken place by 1.30 AU and *then* begins to level off out to the last points beyond 2.0 AU. Finally, our furthest data in 1999 (2.29 AU) are lower by ~4× than the last 2019 point at 2.07 AU.

The cause or causes of these outbound characteristics – a plateau followed by a rapid decrease in production rates, but at different distances from one *apparition* to another – is uncertain. However, one possibility – that this is an artifact due to the use of differing photometer aperture sizes – can readily be eliminated. First, on 2018 September 13 (1.01 AU), the night having the most observational sets, six sizes were utilized spanning a range of more than four times while the derived $Q$(CN) varied by less than 5%. Even $C_2$, the species often showing aperture trends due to its multiple parentage, varied by only 10%. Moreover, within a given apparition, the aperture size we used did not change greatly at the time of these sharp drops. We can also rule out rotational effects, based on our extensive imaging in 2018. While these images will be extensively examined in a future publication, the key aspect relevant here is that the observed CN jet morphology changed extremely little from hour to hour or from night to night, and only gradually varied over the time scale of weeks and months. Discussion of other possibilities that cannot be so readily dismissed are deferred to Section 4.

In terms of temporal coverage, the widest range after our own is that of Moulane et al. (2020) using the same HB filters as ourselves on the pair of robotic TRAPPIST telescopes. With successful gas measurements from -83 days to +100 days, their density of coverage is greater than our own, allowing one to more readily search for either rotational variability or outbursts, though no clear evidence of either is visible in their plots. In addition to their results, they also overlay our early data from 2018 along with our production rates from 1985 and 1998 (see their figure 1). As they note, the agreement is generally excellent, with the largest departures seen after perihelion where our results diverge among the apparitions, and they did not have our late 2018/19 results available; indeed, their continued drop off in $Q$s during this time are consistent with our results just discussed. The other two data sets with smaller but still significant time ranges are both in the IR by Faggi et al. (2019) and Roth et al. (2020), both using the iSHELL spectrograph at NASA's Infrared Telescope Facility (IRTF). With $\Delta T$s from about 6 weeks prior to perihelion to 4 weeks following, both groups detected large drops in production rates for all of their observed parent species, consistent with decreases we and others have observed in the visible and near-UV spectral regimes. Finally, a decrease by a factor of two was also measured in CN, along with a decrease by 1.8× for $C_2$, by Lara et al. (2003) in 1998 during the four weeks surrounding perihelion.

### 3.2 Composition

Based on the overall heliocentric distance behavior of the various gas species as detailed in Section 3.1, it is not surprising that the *relative* production rates between species are roughly constant throughout each apparition. Thus, not only do the mean production rate ratios place G-Z in the strongly carbon-chain depleted category based on the compositional classification by Schleicher & Bair (2016) and recently updated by Bair & Schleicher (2021) but G-Z remains in this grouping throughout each apparition. Looking in more detail, however, reveals some interesting characteristics. While the $Q(C_3)$-to-$Q(C_2)$ ratio shows no trend during an apparition, and each carbon-chain molecule with respect to CN is also nearly static within an apparition, the $Q(C_2)$-to-$Q(CN)$ ratio is clearly *lower throughout 1985 as compared to the later apparitions*, as shown in the top panel of Figure 2. We were initially puzzled by this, until we realized that this is entirely an artifact caused by systematic differences in aperture sizes used in the 1980s as compared to later decades. As previously noted, $C_2$ (and to a lesser degree NH) exhibit aperture trends because the standard 2-generation Haser model does not successfully reproduce the spatial profiles of these multi-generation and/or multi-parent molecules. For $C_2$, smaller apertures yield lower production rates, while the smaller trend for NH is in the opposite direction; such trends with aperture size are confirmed by data obtained within individual nights. Our observations of G-Z in 1985 had a median value of log $\rho$ of 3.76 and only one measurement was made with the log of the projected aperture radius > 4.0 due to most of the telescopes having large f-ratios. By the late 1990's, most observations were made using the 42-in telescope that then had a different, smaller f-ratio secondary, and the corresponding median log aperture radius was 4.5 in 1998/99 and 4.3 in 2018/19. We choose not to attempt a correction for these aperture effects as the specific size of the trend varies somewhat between nights and at different distances, and we never span the full range of aperture sizes at any given time. Rather we note our understanding that it is an artifact of our use of the Haser model and *not* an intrinsic change in G-Z's composition since 1985, even though the specific degree of depletion that we compute is altered by this effect.

[FIGURE 2 HERE: Q-RATIOS VS TIME FOR C2/CN, CN/OH, & NH/OH; 3-PANEL]

A different situation presents itself when comparing the production rate ratios of CN-to-OH, or even NH-to-OH. The $Q(CN)$-to-$Q(OH)$ ratio, shown in the middle panel of Figure 2, decreases significantly from the start of each apparition to the time of perihelion, then levels off, and finally increases by a small amount (and then may decrease at the end of 2018/19, though these points are much more uncertain). While a small component of these variations may be a heliocentric distance effect (possibly an artifact associated with the adopted scalelengths), the asymmetry between the pre- and post-perihelion time frames is real. In particular, the 3× drop in the pre-perihelion time frame for the ratio of CN-to-OH is larger than usual, but changes in this ratio are common for many comets that we have investigated, including Tempel 2 (Knight et al. 2012), 67P/Churyumov-Gerasimenko (Schleicher 2006; see also Gasc et al. 2017), and 2P/Encke (unpublished), where we have attributed this behavior to heterogeneity in composition of the carbon-bearing species to water between different source regions on a comet's nucleus. We assume the same scenario is in effect here for G-Z; the relative abundances among the three carbon-bearing species is the same between the source regions, but the carbon to water ratio differs between sources. Note that this implies the existence of at least two active source regions. Finally, the NH-to-OH ratio is also unusual, in that NH is usually closely tied with OH, but with G-Z it also varies with time, apparently in the opposite manner as the CN-to-OH ratio (though uncertainties are quite large at larger $r_H$). As with nearly all other strongly carbon-chain depleted comets, NH is also depleted as compared either to OH or CN. We will return to these various trends in the production rate ratios in Section 4, but simply note here that we think they are again related to the overall strong seasonal behavior that causes the maximum $Q$s to occur 3-4 weeks prior to perihelion.

As shown above, while some of G-Z's production rate ratios vary by only small amounts throughout its orbit while others vary significantly, none of these alter the resulting compositional classification that G-Z is strongly carbon-chain depleted. The logarithm of the mean ratios for various combinations of species are presented in Table 4, along with the range of values of the other comets in this compositional class and two other relevant classes – moderately carbon-chain depleted and the typical class – for comparison. In our analyses of compositional classifications, we use a variety of diagnostics when determining a comet's classification, including ratio-ratio plots, cluster analysis, and principal component analysis. In the cases of carbon-chain depletion, the key attributes are the ratios of $Q(C_2)$-to-$Q(CN)$, $Q(C_3)$-to-$Q(CN)$, and $Q(NH)$-to-$Q(OH)$. We have recently added five additional years of observations (Bair & Schleicher 2021) to our previously frozen database as of 2016 (Schleicher & Bair 2016). This has resulted in a total of 12 comets (including G-Z) composing the strongly carbon-chain depleted group, having depletion factors of >4.3× in the $C_2$ ratio with respect to CN and >2.7× in the $C_3$ ratio with respect to CN. Nine of the twelve objects also have significant depletion (>2.5×) for the NH to OH abundances. G-Z's values for these three ratios are 4.3×, 5.9×, and 4.6×, respectively, where our fiducials are the mean values for those 102 comets in the typical class. In comparison, the moderately carbon-chain depleted class had eight comets exhibiting some depletion of both $C_2$ and $C_3$ but generally no depletion of NH.

[TABLE 4 HERE: COMPOSITION (Q-RATIOS); DATABASE COMPARISON]

As discussed at the end of Section 3.1, Moulane et al. (2020) also used the HB filter set for their studies (and the same modeling parameters), so that their results are directly comparable to our own. Indeed, their ratios for $C_2$-to-CN and $C_3$-to-CN easily match our own to within the respective uncertainties. However, their mean value for NH-to-OH differs from our own, but they only successfully detected NH during a short interval near perihelion. Given the variation of NH-to-OH we observed over the entire apparition, it is clear that we are in complete agreement once we restrict our results to the same time interval. The other case of using HB filters was by Lara et al. (2003) during four weeks centered on perihelion at the 1998 apparition, and their only ratio, that for $C_2$-to-CN, also matches our corresponding value to within the uncertainties. Remaining in the visible regime but switching to spectroscopic observations, the primary published compositional results are from 1985, including Cochran et al. (1987; 2012), Beaver et al. (1990), and Fink (2009). Direct comparisons are much more difficult either due to the use of differing model parameters and/or measurements of different emission bands, such as a red CN band rather than the violet in the case of Fink. Regardless, there is consensus that $C_2$, $C_3$, NH, and $NH_2$, are all significantly depleted (>2.4×) as compared to CN (and to water, if one of its byproducts was measured).

At longer wavelengths, an ever-increasing number of potential parent species have either been detected or had useful upper-limits determined with each apparition. Mumma et al. (2000) reported that $C_2H_6$ (ethane) was low in 1998, as compared to comets Hale-Bopp (C/1995 O1) and Hyakutake (C/1997 B2) by factors of 2-3×, while Weaver et al. (1999) reported an upper limit for ethane implying an even larger depletion. At the 2005 apparition, DiSanti et al. (2013)

measured a strong depletion of ethane, along with a "mild" depletion of $CH_3OH$ (methanol). Most recently, Roth et al. (2020) again found ethane to be strongly depleted but that $CH_4$ (methane) abundance was consistent with other comets, while Faggi et al. (2019) determined upper-limits for methane, $C_2H_2$ (acetylene), $H_2CO$ (formaldehyde), and $NH_3$ (ammonia). Of these, Faggi et al. report that acetylene and formaldehyde "appear significantly depleted" while ammonia was not sufficiently constrained. They also give an HCN value about 30% less than the average of JFCs, but this production rate is consistent with our CN results; see Section 4 for further discussion.

### 3.3 Water Production and Effective Active Area

As noted in Section 2.3, our OH Haser production rates are readily converted to vectorial equivalent water production rates listed in the final column of Table 3 and shown in the top panel of Figure 3. Comparisons to the values found by other investigators using a variety of techniques, ranging from direct water measurements in the IR to the hydrogen coma with SOHO/SWAN, have already been presented by Combi et al. (2011; 2021) and by Moulane et al. (2020). Direct, detailed comparisons are actually quite difficult due to the sparse cadence of most observations, with only SOHO/SWAN and the TRAPPIST telescopes obtaining data on a fairly regular basis. Our water values in 2018 are intermediate between those from TRAPPIST (systematically lowest) and SOHO/SWAN (see Figure 2 from Moulane et al.), with the differences thought to be associated with technique and modeling, though some variations are evident between apparitions, with the time of peak production ranging between about -40 and -20 days from perihelion. The IR data sets (Faggi et al. 2019; Roth et al. 2020) are much sparser and apparently have greater scatter, possibly due to higher sensitivity to short-term variability either from rotation or small outbursts not evident at larger spatial scales. Overall, though, all data sets are fundamentally in agreement with our findings from 1985 that production rates peak several weeks prior to perihelion and drop rapidly near perihelion. Other specific findings, such as the $r_H$-dependent slope following perihelion, differ between data sets primarily because the range of heliocentric distances used for each calculation on the non-linear production rate curve also differed. In particular, we note that Combi et al. (2021) found significant changes among the apparitions beyond three weeks outbound, consistent with our findings even though some of our data were from different apparitions.

[FIGURE 3 HERE: WATER PRODUCTION & ACTIVE AREA VS TIME; 2 PANEL]

Not surprisingly, given the unusual nature of the production rate curves and peak production taking place several weeks prior to perihelion, the derived effective active areas vary greatly through each apparition, as shown in the bottom panel of Figure 3. Using the standard vaporization model by A'Hearn (2010) based on a correction to the original computations of Cowan & A'Hearn (1979), we obtain a peak value of 14-16 $km^2$, but compute values less than 3 $km^2$ very early and late in the apparition. In fact, only three weeks following perihelion the effective active area is in the range of 4-6 $km^2$ depending on the apparition. If one further assumes an effective nucleus radius of $1.82 \pm 0.05$ km as determined by Pittichová et al. (2008), then the peak active fraction is about 36%, an unusually high value for a Jupiter-family object. Since the general production rate behavior has repeated for the past one-third of a century, it is clear that physical evolution of the active regions must be minimal and that the overall shape of the production rates must be due to seasonal changes as one or more source regions respond to varying amounts of solar illumination. We return to this in Section 4.

*3.4 Dust*

As noted at the beginning of Section 3, the overall behavior of $A(\theta)f\rho$ as a function of time and heliocentric distance was very similar to that of the various gas species, as shown in the lower-left panel of Figure 1. The increased apparent scatter on individual nights, as compared to CN or OH, is due to the near-ubiquitous aperture trend of higher derived $A(\theta)f\rho$ values for smaller apertures. While the cause of this behavior in most comets remains uncertain, and has been attributed to "fading" grains as they darken or shrink in size as the grains coast away from the nucleus (cf. Baum et al. 1992), the steepness of the trend varies among comets. In the case of G-Z this trend with aperture size is relatively moderate, and we have chosen not to make adjustments for aperture size for the same reason we left the $C_2$ production rates "as is" – a single definitive trend could not be determined due to variations between nights and with $r_H$.

Dust measurements over a very similar temporal range in 2018 but with a higher cadence were obtained by Ehlert et al. (2019), to study how material released at this apparition affects the Draconid meteor shower in future years since G-Z is the shower's parent body. Since their CCD extracted fluxes are from a much smaller effective aperture ($10\times10$ arcsec$^2$) than our own data, it was expected that their resulting $A(\theta)f\rho$ would be systematically higher than our own, as was indeed the case. The other large data set was that of Moulane et al. (2020), previously discussed regarding gas productions. Their $A(\theta)f\rho$ extractions used a larger, fixed projected aperture radius of $10^4$ km, and we find excellent agreement with our results when we used similar apertures, and the expected off-sets early and late in the apparition when our projected apertures were larger, with our results systematically lower. However, a different story is found when we compare the resulting $A(\theta)f\rho$ values obtained by Lara et al. (2003) with our own. They made CCD measurements in 1998 during the month surrounding perihelion and determined that $A(\theta)f\rho$ *increased* substantially with increased aperture sizes, quoting values having an average increase of 24% between $\rho$ of $10^4$ km and $1.5\times10^4$ km. In comparison, on our three nights most closely concurrent with theirs, our values are nearly half of theirs, and exhibit aperture trends in the opposite direction. We have no clear explanation for this disagreement.

As a whole, the dust grains exhibit very little color, with a reddening per 1000 Å of only 3% using the green and UV continuum points. This unweighted result, however, is dominated by the scatter in values at larger heliocentric distances due to the intrinsic uncertainties coupled with color being a differential value. If one restricts the analysis to the better determined values, the resulting color increases to 6% per 1000 Å, but there remains considerable spread because there is an unexpectedly large aperture effect on $A(\theta)f\rho$ in the UV filter for small aperture sizes. In fact, the mean value for the green-to-UV color of the good quality data in 1985 is 16% per 1000 Å, while the same color for similar data in the later apparitions is just 1% per 1000 Å, where the apertures were systematically much larger in size as previously discussed in Section 3.1 regarding $C_2$ production rates. While we might hypothesize that the small apertures show greater reddening due to an excess of large, very slow moving grains, if this were true we would expect a change near and following perihelion when the gas flow was highest and larger grains could have been lifted, but we do not see a significant change with $\Delta T$. Whatever the reason, this large trend with aperture size for smaller apertures in the UV explains why the degree of reddening we saw during 1985 is no longer present at later apparitions. Note that a similar strong aperture trend for the UV filter at small projected aperture radii also was measured in Comet 19P/Borrelly (Schleicher et al. 2003) where large, slow moving grains were invoked for multiple reasons.

Though never approaching 0°, the phase angle of G-Z varied over a wide range, 21° to 78°, encompassing a range of phase angle adjustments spanning nearly a factor of 1.5× (see Table 1). We, therefore, have applied our now-standard phase angle correction (cf. Schleicher & Bair 2011) to compute normalized $A(\theta)f\rho$ values, i.e. $A(0°)f\rho$, and these values for the green continuum are plotted in the top panel of Figure 4. (Note that phase corrections do not alter the previous discussion regarding aperture effects or colors since those characteristics are measured on individual nights and thus the phase adjustments would have been identical and cancelled out.) The bulk $r_H$-dependence for $A(0°)f\rho$, computed in a manner consistent with the gas values reported early in Section 3.1, had a log-log slope of -3.5 ± 0.3, intermediate to the carbon-bearing species and OH. This is somewhat unusual in that, especially following perihelion, the $r_H$-dependence of dust is usually significantly shallower than *any* gas species. Finally, we compute the dust-to-gas ratio, based on the log $A(0°)f\rho$ – log $Q$(OH), and these results are shown in the bottom panel of Figure 4. (We use OH rather than $H_2O$ since the definition of $Af\rho$ is a better match to the Haser model.) As is evident in the figure, the ratio is fairly constant until late in each apparition when values increase by moderate amounts. The mean value, overall, is -25.46 cm s molecule$^{-1}$, about one-third of the phase adjusted values we find for spacecraft targets 1P/Halley or 67P/Churyumov-Gerasimenko (hereafter C-G), but in the mid-range for comets overall in our database.

[FIGURE 4 HERE: DUST (PHASE CORRECTED) VS TIME; DUST/GAS VS TIME; 2 PANEL]

## 4. DISCUSSION AND SUMMARY

Nearly a century ago, 21P/Giacobini-Zinner was the first comet reported to exhibit an unusual spectrum with the emission of certain common species, in particular $C_2$, much fainter than usual when compared with CN. In 1985, our measurements of OH in addition to the minor species provided the evidence required to confirm that $C_2$ (and $C_3$) must be strongly depleted rather than CN being strongly enhanced. In addition, we determined that NH was also depleted, while others found that $NH_2$ was similarly depleted and confirmed various abundance ratios. By the time of our first large database paper, we had discovered several other comets having varying degrees of depletion and thus G-Z became the proto-type of the carbon-chain depleted compositional class. In particular, most but not all were also depleted in $C_3$, while many in the group were also depleted in NH. An additional quarter century of observations greatly enlarged our sample size from 41 well-determined comets to 135, and as of mid-2021 we formally divided those comets exhibiting carbon depletion into three classes: strongly carbon-chain depleted (12 objects); moderately carbon-chain depleted (8); and $C_2$ but *not* $C_3$ depleted (6). Note that the division between the two carbon-chain depleted classes is somewhat arbitrary as there is a continuum regarding the degree of depletion among these 20 comets, and our cluster analysis combined with numerous ratio-ratio plots provided the dividing line. Based only on $C_2$, G-Z falls at the split, with the somewhat greater $C_3$ depletion pushing G-Z into the "strongly" side. The other ratio of note is that of NH-to-OH. Of these three carbon depleted classes, only comets in the strongly carbon-chain depleted class usually (but not all cases) also exhibit NH depletion, as does G-Z. Overall, we find that Giacobini-Zinner continues to be an excellent representative of carbon-chain depleted comets, and in particular of the strongly carbon-chain depleted classification.

All studies of G-Z's chemical composition based on observations in the visible and near-UV give excellent agreement on a number of fronts. The degree of depletions is consistent and, when allowing for differences in reduction parameters or modelling limitations, there is *no* evidence for significant changes in depletions as a function of time or heliocentric distance, and apparently the degree of depletions is unchanged from 1985 to 2018, and probably going back an entire century. Thus, G-Z's abundance ratios are very stable. This in turn suggests that G-Z's depleted abundances are *not* associated with an unusual, active component of the body having a different composition from the currently inert remainder of the nucleus. Preliminary results from our imaging campaign suggests that two rather than one source regions are producing gas jets. If confirmed with modeling, this would further support a homogeneous composition across the surface of the nucleus among the carbon-bearing species, though we noted that the changing CN-to-OH ratio with orbital position implies some compositional differences between sources. Additional evidence comes from another strongly carbon-chain depleted object, 73P/Schwassmann-Wachmann 3, which partially fragmented, thereby yielding a high active fraction, and the freshly exposed ice of several fragments exhibited the same degree of depletion as the original body (Schleicher & Bair 2011). Not only did these results directly imply that the entire nucleus has a homogeneous composition, but that the depleted composition was *not* caused by thermal or other evolutionary process, and was instead associated with its origin. We conclude the same is true for G-Z.

Considerable effort has been made to improve understanding of a variety of parent molecules accessible in the IR and millimeter regimes, and while G-Z was never sufficiently bright to detect many of these species, successful measurements or constraining upper-limits were obtained, particularly during the 2018/19 apparition. In addition to $H_2O$, the known parent of OH, hydrogen cyanide (HCN) was first detected in G-Z on September 8 & 9 (Faggi et al. 2019) at production rates consistent with it being the dominant if not sole parent of CN [unpublished radio measurements of HCN may also exist; Cordiner et al. 2019]. Unfortunately, upper-limits on $NH_3$ are not constraining, though there is little doubt that ammonia must be depleted given that NH and $NH_2$ are both significantly depleted by quite similar amounts. Of the other potential parent species of those molecules we and others have observed in the visible regime, the most notable are acetylene ($C_2H_2$) and ethane ($C_2H_6$), though we note that $C_3$ is obviously an additional parent of $C_2$. Ethane has now been reported as strongly depleted at both viable apparitions, possibly even at a greater depletion rate than $C_2$, but this is somewhat uncertain since "normal" values are computed with varying sets of comets among the research groups. The most constraining upper-limits for acetylene suggest that it is depleted by a similar amount as ethane. Finally, though not a parent of one of "our" molecules, we note that the upper-limit of $H_2CO$ (formaldehyde) by Faggi et al. (2019) implies that this species is also strongly depleted, while they report that $CH_3OH$ (methanol) is near normal but exhibits an apparent seasonal variation when compared to water.

All evidence continues to indicate that G-Z, and indeed *all* of the comets exhibiting carbon-chain depletion, have this composition due to the conditions in the early solar nebula when and where they formed, rather than due to recent evolution when they subsequently arrived in the inner solar system. We continue to assume that the temperature of condensation out of the nebula into cometesimals was the key factor, and the range of carbon-chain depletions was associated with a range of heliocentric distances and thus temperature from which many JFCs were formed. The actual location for this and whether it was closer or further from the Sun than for non-depleted objects remains both complicated and controversial. While it had been generally thought that Oort Cloud objects originated closer to the Sun and since very few of these objects are carbon-chain depleted, we tentatively concluded that the carbon-chain depletion increases with greater distances and lower temperatures (A'Hearn et al 1995). Other evidence suggests that G-Z formed in a warmer region (Ootsubo et al. 2020), and that perhaps JFCs as a class formed at somewhat warmer temperatures based on the relative abundances of $H_2O$, CO, and $CO_2$ (A'Hearn et al. 2012). However, 67P/C-G is a member of the moderately carbon-chain depleted class (Schleicher 2006), and considerable evidence from the Rosetta mission indicates that C-G formed at quite low temperatures (cf. Rubin et al. 2015; 2018), possibly implying that carbon-chain depleted comets, including G-Z, formed further out than the large majority of comets. We will explore these issues in more detail in our upcoming database paper (Schleicher & Bair, in preparation).

The cause of the NH and $NH_2$ depletion in G-Z is more uncertain. Since fewer carbon-chain objects also exhibit NH depletion, the simple answer is that ammonia depletion only takes place at even colder condensation temperatures than for carbon-chain parent species. However, there is also a group of comets in our database that show significant depletion of NH but are not carbon-chain depleted (Schleicher & Bair 2016), suggesting an alternative method of ammonia depletion. Whether these distinct cases and groupings have a common cause or NH depletion occurs through differing paths remains unclear.

The very large seasonal changes observed in G-Z in 1985 continue to be present 33 years later and, except for the detailed behavior between ~30 and ~80 days following perihelion, the shape of the production rate curves for every gas species and the dust are essentially the same. Production peaks 3-5 weeks prior to perihelion, then plummets near perihelion, before diverging outbound among the three apparitions, but with all species showing the same behavior *within* an apparition. This is consistent with our narrowband imaging in that while morphological changes occurred during the 2018/19 apparition, all gas species exhibited the same features at any one time. Our suspicion is that the outbound changes in the production rate curve from one apparition to the next is most likely due to a long-term precession of the pole. Even a moderate change could significantly alter the amount of available sunlight if a source region was illuminated obliquely due to the Sun remaining near the local horizon – a likely condition given the overall low production rates after perihelion. Note that Sekanina (1985) developed an extensive model of precession during the 1900s based on non-gravitational perturbations in G-Z's orbital motion, but as Schleicher et al. (1987) showed, Sekanina's predictions for the seasonal behavior at 1985 apparition, including having peak production occurring several weeks *following* perihelion, proved very different from reality. Though his precession model proved incorrect, we think precession remains the most plausible explanation for only a portion of the orbit showing substantial changes in outgassing over only a few orbits. In particular, even a small amount of precession could produce a large effect when the Sun is low in the local sky, i.e. post-perihelion, but have a minimal effect when the Sun is high overhead, thereby explaining the lack of significant changes in the seasonal behavior prior to perihelion. This, of course, also assumes that the primary source region is itself located at a high latitude, thereby permitting a large seasonal effect if the rotation axis has a sufficiently large obliquity. Finally, the basic fact that peak production takes place well before perihelion strongly suggests that a localized source region saw maximum solar illumination even earlier, with the sub-solar latitude already decreasing at the time of the observed production rate maximum. The trade-off of solar intensity due to heliocentric distance as compared to the Sun's changing zenith distance results in our estimate that the Sun's sub-solar latitude reached a peak several weeks prior to the production rate maximum, i.e. perhaps as much as 7-9 weeks before perihelion. Our planned jet modeling will test this prediction, as well as our expectation that a secondary source region, having a different relative abundance of OH as compared to all of the carbon-bearing species, is located at a quite different latitude than the primary source to explain this ratio's large seasonal variation.

We noted in Section 3.3 the unusually large effective active area and the correspondingly large fractional active area, ~36%, that we computed to produce the measured peak water production, based on the standard water vaporization model. This standard model, assuming an isothermal sphere, is routinely applied since it makes the fewest assumptions regarding a comet's nucleus and provides an upper-limit on the required area. However, it is apparent that the standard model is inappropriate in the case of G-Z. First, if more than a third of the total surface were active, it would be nearly impossible to get such a large pre-/post-perihelion offset in production rates as is observed. Second, the associated minimum active fractions of a few percent late in the apparition must also be explained. If, as was just discussed, the primary source region is located at a near-polar location and the Sun was high in its sky before perihelion, then the standard isothermal sphere should be replaced by the pole-on vaporization model, which is approximately four times more efficient (A'Hearn 2010), and would only require the source to be ~9% of the total nucleus. A secondary source, located far from the pole, would be expected to only modestly modify this result. We also note, based on our OH imaging, at least a small component of water appears to be released from the surface in the form of icy grains, potentially further reducing the required active fraction. Thus, G-Z's active fraction is much closer to the median value of ~10% for JFCs (Bair & Schleicher 2021) than it first appeared. Note that the peak in production rates for 67P/C-G was also 4-5 weeks away from perihelion, though with the opposite sense (Schleicher 2006), and this was associated with strong polar illumination along with evidence of material falling back onto the other hemisphere (Keller et al. 2017); a similar mechanism of dust transfer might be a contributor here in the case of G-Z.

Finally, we note that both the dust properties in G-Z as well as the resulting dust-to-gas ratio are quite unremarkable. Our measure of the dust, $A(\theta)f\rho$, shows the usual decrease in values with increasing aperture size, directly indicating some type of fading grains, just as we find in most comets. The dust-to-gas ratio is in the mid-range of values we've previously found for JFCs, though the full range covers two orders of magnitude. In general, the G-Z's dust grains are slightly red in color, but the color is more pronounced in the ultraviolet at the smallest apertures, another characteristic we have seen in other comets.

Overall, Comet 21P/Giacobini-Zinner is arguably now the best studied object of the strongly carbon-chain depleted compositional class, with the possible exception of 73P/Schwassmann-Wachmann 3. In addition to being a key member of this class, G-Z has continued to reveal many properties unique to itself. Our own studies will continue with analyses of the gas morphology revealed from our extensive imaging campaign and eventual modeling, with the expectation of revealing the causes of G-Z's near unique seasonal behavior.

## ACKNOWLEDGEMENTS

We gratefully acknowledge Robert Millis for the acquisition of many of the original 1985 observations, along with the assistance of Peter Birch. In 1998/99, Tony Farnham obtained a number of the observations while P. Birch acquired three nights of data at Perth Observatory. We particularly thank Allison Bair for assisting with photometric analyses and the construction of tables, along with numerous discussions regarding comparison with our photometric database and the compositional classifications. This work was supported by NASA Solar System Observations Program grant 80NSSC18K0856, and predecessor grants.

Facilities: Lowell Observatory 42 inch (1.1 m) John S. Hall Telescope; Lowell Observatory 72 inch (1.8 m) Perkins Telescope; Lowell Observatory 31 inch (0.8 m) telescope (LO:0.8m); Perth Observatory 24 inch (0.6 m) Lowell Telescope.

Software: Data Desk 6.3 by Data Description, Inc.

ORCID iD:  David Schleicher  https://orcid.org/0000-0003-1412-2511

## REFERENCES

A'Hearn, M. F. 1991. Photometry and polarimetry network. In The Comet Halley Archive Summary Volume (Z. Sekanina, Ed.), pp. 193–235. Jet Propulsion Laboratory, Pasadena.

A'Hearn, M. F. 2010, PDS/SBN water vaporization model; [unclear how to reference]. https://pds-smallbodies.astro.umd.edu/tools/ma-evap/index.shtml

A'Hearn, M. F., Feaga, L. M., Keller, H. U., et al. 2012, ApJ 758, 29.

A'Hearn, M. F., Millis, R. L., Schleicher, D. G., Osip, D. J., & Birch, P. V., 1995. Icarus 118, 223.

A'Hearn, M. F., Schleicher, D. G., Feldman, P. D., Millis, R. L., & Thompson, D. T. 1984, AJ 89, 579.

Bair, A. N., & Schleicher, D. G. 2021, AAS/DPS Meeting, 53, 210.03.

Baum, W.A., Kreidl, T.J., Schleicher, D. G., 1992, AJ 104, 1216.

Beaver, J. E., Wagner, R. M., Schleicher, D. G., & Lutz, B. L. 1990, ApJ 360, 696.

Bobrovnikoff, N. T. 1927, ApJ 66, 439.

Cochran, A. L., & Barker, E. S. 1987, ApJ 93, 239.

Cochran, A. L. & Schleicher, D. G. 1993, Icarus 105, 235.

Cochran, A. L., Nelson, T., and McKay, A. J. 2020, PSJ 1: 71.

Combi, M. R., Lee, Y., Patel, T. S., Mäkinen, J. T. T., Bertaux, J.-L., Quémerais, E. 2011, AJ 141, 128.

Combi, M. R., Mäkinen, T., Bertaux, J.-L., Quémerais, E., Ferron, S., & Coronel, R. 2021, Icarus 357, 114242.

Cordiner, M., Biver, N., Milam, S., et al. 2019, EPSC-DPS Joint Meeting 2019, 1131.

Cowan, J. J., & A'Hearn, M. F. 1979, M&P 21, 155.

DiSanti, M. A., Bonev, B. P., Villanueva, G. L., & Mumma, M. J. 2013, ApJ 763, 1.

Ehlert, S., Moticska, N., & Egal, A. 2019, AJ 158: 7.

Faggi, S., Mumma, M. J., Villanueva, G. L., Paganini, L., & Lippi, M. 2019, AJ 158: 254.

Farnham, T. L., Schleicher, D. G., & A'Hearn, M. F. 2000, Icarus 147, 180.

Farnham, T. L., & Schleicher, D. G. 2005, Icarus, 173, 533.

Gasc, S., Altwegg, K., Balsiger, H., et al. 2017, MNRAS 463, S108.

Haser, L. 1957 Bull. Cl. Sci. Acad. R. Belg. 43, 740.

Keller, H. U., Mottola, S., Hviid, S. F., et al. 2017, MNRAS, 469, S357.

Knight, M. M., & Schleicher, D. G. 2013, Icarus 222, 691.

Knight, M. M., Schleicher, D. G., Farnham, T. L., Schwieterman, E. W., & Christensen, S. R. 2012, AJ 144, 153.

Lara, L.-M., Licandro, J., Oscoz, A., & Motta, V. 2003, A&A 399, 763.

McFadden, L. A., A'Hearn, M. F., Feldman, P. D., et al. 1987, Icarus 69, 329.

Meier, R., Wellnitz, D., Kim, S. J., & A'Hearn, M. F. 1998, Icarus, 136, 268.

Mianes, P., Grudzinska, S., and Stawikowski, A. 1960, Ann. Astrophys. 23, 788.

Moulane, Y., Jehin, E., Rousselot, P., et al. 2020, A&A 640, A54.

Mumma, M. J., DiSanti, M. A., Dello Russo, N., Magee-Sauer, K., & Rettig, T. W. 2000, ApJ 531, L155.

Ootsubo, T., Kawakita, H., Shinnaka, Y., Watanabe, J., & Honda, M. 2020, Icarus 338, 113450.

Pittichová, J., Woodward, C. E., Kelley, M. S., & Reach, W. T. 2008, AJ 136, 1127.

Roth, N. X., Gibb, E. L., Bonev, B. P., et al. 2020, AJ 159: 42.

Rubin, M., Altwegg, K., Balsiger, H., et al. 2015, Science 348, 6231.

Rubin, M., Altwegg, K., Balsiger, H., et al. 2018, Sci. Adv. 4, eaar6297.

Schleicher, D. G. 2006, Icarus 181, 442.

Schleicher, D. G. 2010, AJ 140, 973.

Schleicher, D. G., & A'Hearn, M. F. 1988, ApJ 331, 1058.

Schleicher, D. G. & Bair, A. N. 2011, AJ 141, 177.

Schleicher, D. G., & Bair, A. N. 2016, AAS/DPS Meeting, 48, 308.04.

Schleicher, D. G., Millis, R. L., & Birch, P. V. 1998, Icarus, 132, 397.

Schleicher, D. G., Millis, R. L., & Birch, P. V. 1987, A&A 187, 531.

Schleicher, D. G. & Osip, D. J. 2002, Icarus 159, 210.

Schleicher, D. G., Woodney, L. M., & Millis, R. L. 2003, Icarus 162, 415.

Weaver, H. A., Chin, G., Bockel´ee-Morvan. D. et al. 1999, Icarus 142, 482.

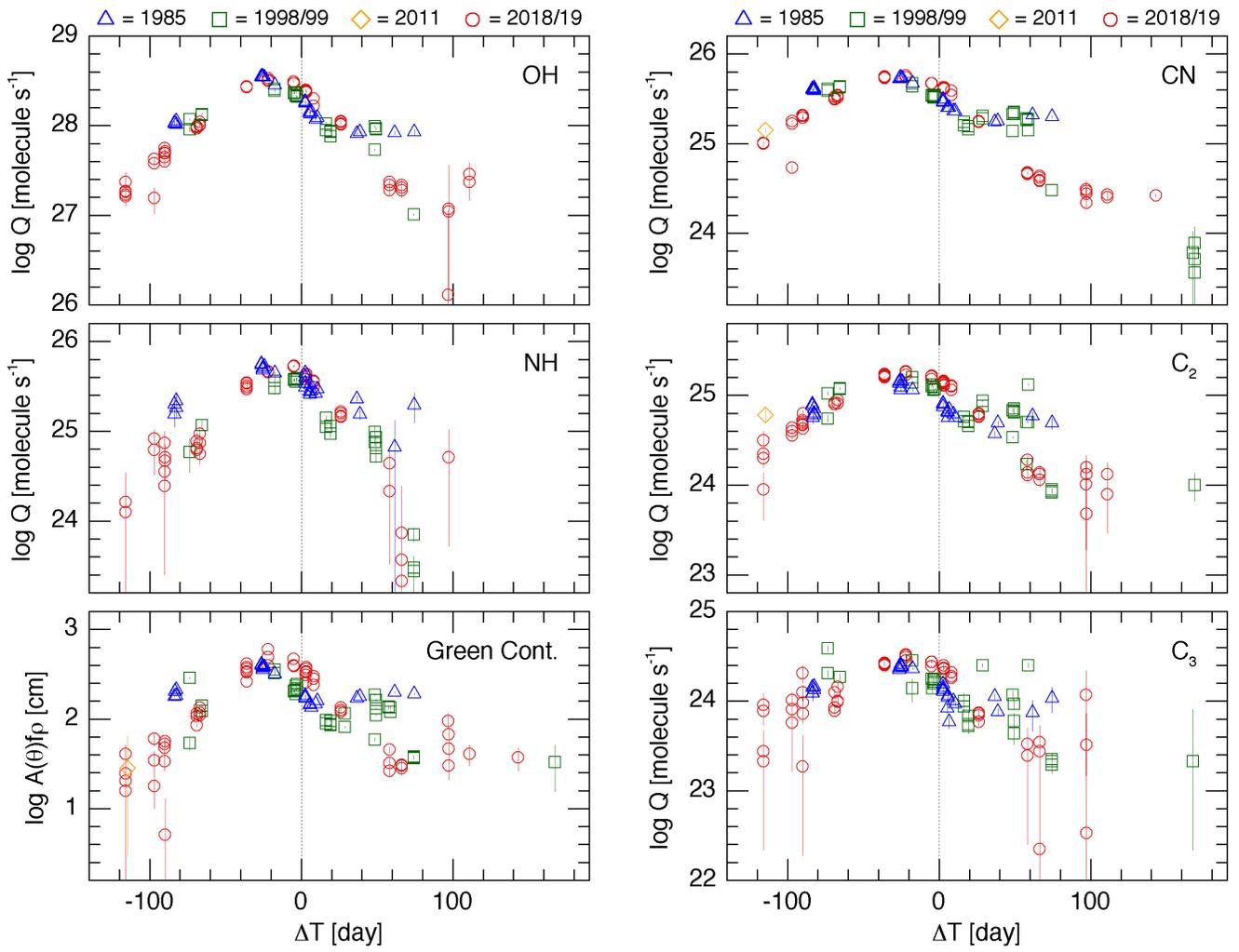

Figure 1

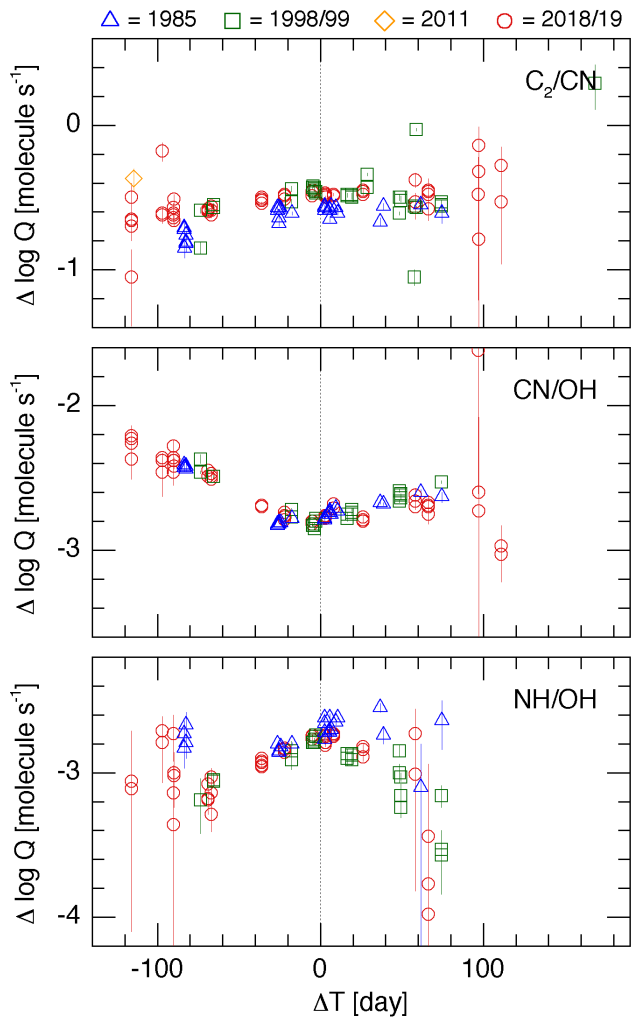

Figure 2

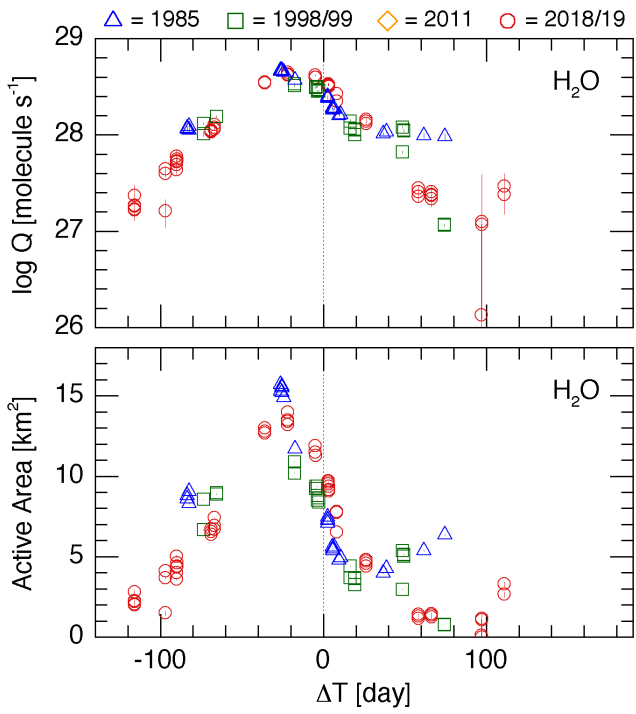

Figure 3

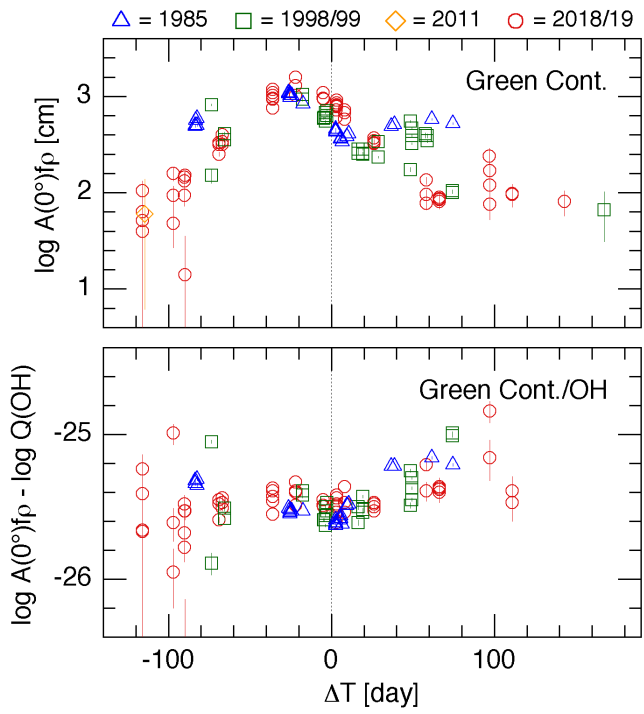

Figure 4

Table 1. Photometry observing circumstances and fluorescence efficiencies for comet 21P/Giacobini-Zinner. [a]

| UT Date | ΔT | $r_H$ | Δ | Phase | Phase | $\dot{r}_H$ | log $L/N$ (erg s$^{-1}$ molecule$^{-1}$) | | | Teles. [c] |
|---|---|---|---|---|---|---|---|---|---|---|
| | (day) | (AU) | (AU) | (°) | Adj. [b] | (km s$^{-1}$) | OH | NH | CN | |
| 1985 Jun 15.4 | −83.4 | 1.515 | 1.001 | 41.7 | 0.446 | −15.5 | −15.102 | −13.489 | −12.815 | L72 |
| 1985 Jun 16.4 | −82.4 | 1.506 | 0.991 | 42.0 | 0.447 | −15.5 | −15.097 | −13.484 | −12.810 | L72 |
| 1985 Aug 11.4 | −26.4 | 1.091 | 0.541 | 67.3 | 0.437 | −8.1 | −14.842 | −13.194 | −12.489 | L72 |
| 1985 Aug 12.4 | −25.4 | 1.086 | 0.535 | 67.8 | 0.435 | −7.8 | −14.848 | −13.190 | −12.488 | L72 |
| 1985 Aug 13.4 | −24.3 | 1.082 | 0.529 | 68.3 | 0.432 | −7.5 | −14.851 | −13.186 | −12.487 | L72 |
| 1985 Aug 20.4 | −17.4 | 1.055 | 0.494 | 71.3 | 0.417 | −5.5 | −14.810 | −13.170 | −12.495 | L72 |
| 1985 Sep 9.4 | +2.7 | 1.029 | 0.454 | 74.4 | 0.398 | +1.2 | −14.824 | −13.192 | −12.559 | L42 |
| 1985 Sep 12.4 | +5.7 | 1.032 | 0.457 | 73.9 | 0.401 | +2.2 | −14.770 | −13.165 | −12.495 | L31 |
| 1985 Sep 13.5 | +6.7 | 1.034 | 0.459 | 73.7 | 0.403 | +2.5 | −14.750 | −13.159 | −12.476 | L31 |
| 1985 Sep 16.4 | +9.6 | 1.039 | 0.465 | 72.9 | 0.407 | +3.5 | −14.693 | −13.136 | −12.421 | L72 |
| 1985 Sep 17.5 | +10.7 | 1.041 | 0.468 | 72.5 | 0.410 | +3.8 | −14.678 | −13.128 | −12.409 | L72 |
| 1985 Oct 13.5 | +36.7 | 1.155 | 0.597 | 59.7 | 0.463 | +10.8 | −14.567 | −13.205 | −12.456 | L72 |
| 1985 Oct 15.5 | +38.7 | 1.168 | 0.609 | 58.6 | 0.465 | +11.2 | −14.547 | −13.221 | −12.472 | L72 |
| 1985 Nov 7.5 | +61.7 | 1.339 | 0.754 | 46.9 | 0.463 | +14.3 | −14.487 | −13.377 | −12.622 | L42 |
| 1985 Nov 20.5 | +74.7 | 1.450 | 0.833 | 41.1 | 0.444 | +15.2 | −14.551 | −13.453 | −12.686 | L42 |
| 1998 Sep 9.2 | −73.6 | 1.435 | 1.184 | 43.9 | 0.454 | −15.0 | −15.011 | −13.442 | −12.770 | L42 |
| 1998 Sep 17.1 | −65.6 | 1.367 | 1.152 | 46.1 | 0.461 | −14.4 | −14.917 | −13.407 | −12.726 | L42 |
| 1998 Nov 4.1 | −17.7 | 1.063 | 0.908 | 59.8 | 0.462 | −5.7 | −14.818 | −13.173 | −12.498 | L31 |
| 1998 Nov 17.1 | −4.7 | 1.036 | 0.865 | 61.9 | 0.457 | −1.5 | −14.830 | −13.251 | −12.590 | L42 |
| 1998 Nov 18.1 | −3.6 | 1.035 | 0.863 | 62.0 | 0.457 | −1.2 | −14.836 | −13.253 | −12.599 | L42 |
| 1998 Nov 19.1 | −2.7 | 1.034 | 0.862 | 62.1 | 0.456 | −0.8 | −14.848 | −13.250 | −12.607 | L42 |
| 1998 Dec 8.1 | +16.4 | 1.060 | 0.869 | 60.5 | 0.461 | +5.4 | −14.633 | −13.117 | −12.379 | L31 |
| 1998 Dec 11.1 | +19.3 | 1.070 | 0.878 | 59.8 | 0.462 | +6.3 | −14.629 | −13.114 | −12.380 | L31 |
| 1998 Dec 20.5 | +28.8 | 1.111 | 0.919 | 57.0 | 0.467 | +8.8 | −14.644 | −13.146 | −12.412 | P24 |
| 1999 Jan 9.1 | +48.4 | 1.234 | 1.059 | 50.1 | 0.469 | +12.6 | −14.491 | −13.290 | −12.547 | L42 |
| 1999 Jan 10.1 | +49.4 | 1.241 | 1.068 | 49.7 | 0.468 | +12.7 | −14.489 | −13.296 | −12.553 | L42 |
| 1999 Jan 18.5 | +57.8 | 1.306 | 1.148 | 46.7 | 0.463 | +13.8 | −14.480 | −13.352 | −12.604 | P24 |
| 1999 Jan 19.5 | +58.8 | 1.314 | 1.158 | 46.4 | 0.462 | +13.9 | −14.481 | −13.358 | −12.609 | P24 |
| 1999 Feb 4.1 | +74.4 | 1.444 | 1.330 | 41.4 | 0.445 | +15.0 | −14.547 | −13.447 | −12.682 | L72 |
| 1999 May 8.2 | +167.4 | 2.280 | 2.734 | 20.8 | 0.299 | +15.0 | −14.943 | −13.842 | −13.100 | L72 |
| 1999 May 9.2 | +168.4 | 2.289 | 2.750 | 20.6 | 0.297 | +15.0 | −14.947 | −13.845 | −13.103 | L72 |
| 2011 Oct 20.1 | −114.6 | 1.806 | 2.324 | 24.0 | 0.331 | −15.9 | −15.288 | −13.642 | −12.971 | L42 |
| 2018 May 17.3 | −116.0 | 1.818 | 1.342 | 33.2 | 0.404 | −16.0 | −15.302 | −13.648 | −12.979 | L42 |
| 2018 Jun 5.3 | −97.0 | 1.642 | 1.108 | 37.3 | 0.428 | −16.0 | −15.213 | −13.562 | −12.886 | L42 |
| 2018 Jun 12.2 | −90.0 | 1.578 | 1.032 | 39.2 | 0.437 | −15.9 | −15.171 | −13.527 | −12.851 | L42 |
| 2018 Jul 3.2 | −69.1 | 1.389 | 0.829 | 46.6 | 0.463 | −15.1 | −14.991 | −13.415 | −12.740 | L42 |
| 2018 Jul 5.2 | −67.1 | 1.372 | 0.812 | 47.4 | 0.464 | −14.9 | −14.963 | −13.405 | −12.728 | L42 |
| 2018 Aug 5.2 | −36.0 | 1.135 | 0.561 | 63.2 | 0.453 | −10.8 | −14.666 | −13.257 | −12.558 | L42 |
| 2018 Aug 19.4 | −21.9 | 1.060 | 0.465 | 71.3 | 0.417 | −7.3 | −14.836 | −13.169 | −12.471 | L42 |
| 2018 Sep 5.3 | −4.9 | 1.015 | 0.396 | 77.7 | 0.376 | −1.7 | −14.807 | −13.231 | −12.565 | L42 |
| 2018 Sep 13.4 | +3.1 | 1.014 | 0.393 | 77.7 | 0.376 | +1.1 | −14.815 | −13.182 | −12.551 | L42 |
| 2018 Sep 18.4 | +8.1 | 1.020 | 0.402 | 76.5 | 0.384 | +2.9 | −14.712 | −13.139 | −12.438 | L42 |
| 2018 Oct 6.4 | +26.1 | 1.080 | 0.486 | 67.5 | 0.436 | +8.4 | −14.627 | −13.119 | −12.387 | L42 |
| 2018 Nov 7.4 | +58.1 | 1.297 | 0.704 | 49.1 | 0.468 | +14.2 | −14.462 | −13.349 | −12.593 | L42 |
| 2018 Nov 15.4 | +66.1 | 1.364 | 0.758 | 45.2 | 0.459 | +14.9 | −14.496 | −13.398 | −12.633 | L42 |

Table 1—Continued

| UT Date | ΔT | $r_{\rm H}$ | Δ | Phase | Phase | $\dot{r}_{\rm H}$ | log $L/N$ (erg s$^{-1}$ molecule$^{-1}$) | | | Teles. [c] |
|---|---|---|---|---|---|---|---|---|---|---|
| | (day) | (AU) | (AU) | (°) | Adj. [b] | (km s$^{-1}$) | OH | NH | CN | |
| 2018 Dec 16.4 | +97.1 | 1.643 | 0.964 | 32.9 | 0.403 | +16.0 | −14.676 | −13.569 | −12.801 | L42 |
| 2018 Dec 30.3 | +111.0 | 1.772 | 1.071 | 29.0 | 0.375 | +16.1 | −14.745 | −13.635 | −12.870 | L42 |
| 2019 Jan 31.3 | +143.0 | 2.066 | 1.398 | 24.6 | 0.337 | +15.7 | −14.866 | −13.764 | −13.007 | L42 |

[a]All parameters are given for the midpoint of each night's observations.

[b]Adjustment to 0° solar phase angle to log($A(\theta)f\rho$) values based on assumed phase function (see text).

[c]Telescope ID: L72 = Lowell 72-inch (1.8-m); L42 = Lowell 42-inch (1.1-m); L31 = Lowell 31-inch (0.8-m); P24 = Perth Observatory 24-inch (0.6-m).

Table 2.   Photometric fluxes and aperture abundances for comet 21P/Giacobini-Zinner.

| UT Date | Aperture | | log Emission Band Flux[a] | | | | | log Continuum Flux[a] | | | log $M(\rho)$[a] | | | | |
|---|---|---|---|---|---|---|---|---|---|---|---|---|---|---|---|
| | Size | log $\rho$ | | | (erg cm$^{-2}$ s$^{-1}$) | | | | (erg cm$^{-2}$ s$^{-1}$ Å$^{-1}$) | | | | (molecule) | | |
| | (arcsec) | (km) | OH | NH | CN | C$_3$ | C$_2$ | UV | Blue | Green | OH | NH | CN | C$_3$ | C$_2$ |
| 1985 Jun 15.3 | 20.2 | 3.87 | −11.33 | −12.79 | −11.25 | −11.98 | −12.06 | −14.22 | ... | −13.91 | 31.23 | 28.15 | 29.01 | 27.83 | 28.10 |
| 1985 Jun 15.4 | 14.1 | 3.71 | ... | ... | −11.52 | −12.19 | −12.47 | −14.36 | ... | −14.02 | ... | ... | 28.75 | 27.62 | 27.69 |
| 1985 Jun 16.3 | 20.2 | 3.86 | −11.33 | −12.64 | −11.26 | −11.90 | −12.17 | −14.27 | ... | −13.90 | 31.21 | 28.29 | 28.99 | 27.90 | 27.98 |
| 1985 Jun 16.4 | 20.2 | 3.86 | −11.29 | −12.71 | −11.24 | −11.93 | −12.09 | −14.25 | ... | −13.90 | 31.25 | 28.21 | 29.01 | 27.87 | 28.05 |
| 1985 Jun 16.4 | 14.1 | 3.70 | ... | ... | −11.49 | −12.14 | −12.41 | −14.33 | ... | −13.99 | ... | ... | 28.76 | 27.65 | 27.74 |
| 1985 Aug 11.4 | 28.5 | 3.75 | −10.00 | −11.40 | −10.29 | −10.97 | −11.00 | −13.14 | ... | −12.86 | 31.75 | 28.71 | 29.12 | 28.02 | 28.34 |
| 1985 Aug 11.4 | 20.2 | 3.60 | −10.26 | −11.67 | −10.54 | −11.14 | −11.26 | −13.31 | ... | −13.02 | 31.49 | 28.44 | 28.87 | 27.85 | 28.08 |
| 1985 Aug 12.4 | 28.5 | 3.74 | −10.01 | −11.44 | −10.28 | −10.92 | −11.04 | −13.21 | ... | −12.89 | 31.75 | 28.65 | 29.12 | 28.06 | 28.28 |
| 1985 Aug 12.4 | 40.1 | 3.89 | −9.77 | −11.21 | −10.04 | −10.73 | −10.74 | −13.08 | ... | −12.76 | 31.98 | 28.88 | 29.36 | 28.25 | 28.58 |
| 1985 Aug 12.5 | 14.1 | 3.44 | −10.52 | ... | −10.77 | −11.31 | −11.58 | −13.52 | ... | −13.17 | 31.23 | ... | 28.63 | 27.67 | 27.74 |
| 1985 Aug 13.4 | 28.5 | 3.74 | −10.02 | −11.42 | −10.27 | −10.92 | −10.98 | −13.18 | ... | −12.87 | 31.73 | 28.66 | 29.11 | 28.04 | 28.33 |
| 1985 Aug 20.4 | 28.5 | 3.71 | −10.04 | −11.45 | −10.31 | −10.89 | −11.02 | −13.24 | ... | −12.89 | 31.61 | 28.56 | 29.02 | 27.99 | 28.21 |
| 1985 Sep  9.4 | 35.3 | 3.76 | −10.07 | −11.41 | −10.41 | −10.97 | −11.01 | −13.33 | ... | −12.98 | 31.52 | 28.55 | 28.91 | 27.81 | 28.12 |
| 1985 Sep  9.4 | 35.3 | 3.76 | −10.06 | −11.44 | −10.40 | −10.95 | −10.97 | −13.30 | ... | −13.00 | 31.53 | 28.51 | 28.92 | 27.84 | 28.16 |
| 1985 Sep  9.5 | 24.8 | 3.61 | −10.30 | −11.54 | −10.63 | −11.13 | −11.24 | −13.46 | ... | −13.14 | 31.28 | 28.41 | 28.69 | 27.66 | 27.90 |
| 1985 Sep  9.5 | 49.8 | 3.91 | −9.82 | −11.09 | −10.17 | −10.72 | −10.75 | −13.20 | ... | −12.86 | 31.77 | 28.87 | 29.16 | 28.06 | 28.39 |
| 1985 Sep 12.4 | 57.6 | 3.98 | −9.78 | −11.12 | −10.08 | −10.94 | −10.74 | −13.16 | ... | −12.88 | 31.76 | 28.82 | 29.18 | 27.85 | 28.41 |
| 1985 Sep 12.4 | 57.6 | 3.98 | −9.79 | −11.03 | −10.08 | −10.80 | −10.75 | −13.20 | ... | −12.88 | 31.75 | 28.90 | 29.18 | 28.00 | 28.39 |
| 1985 Sep 12.5 | 40.4 | 3.83 | −10.04 | −11.40 | −10.32 | −10.99 | −11.04 | −13.35 | ... | −13.02 | 31.49 | 28.53 | 28.94 | 27.81 | 28.10 |
| 1985 Sep 13.5 | 57.6 | 3.98 | −9.76 | −11.10 | −10.07 | −11.09 | −10.72 | −13.10 | ... | −12.91 | 31.76 | 28.83 | 29.18 | 27.71 | 28.43 |
| 1985 Sep 16.4 | 28.5 | 3.68 | −10.28 | −11.63 | −10.53 | −11.22 | −11.26 | −13.51 | ... | −13.19 | 31.20 | 28.29 | 28.67 | 27.60 | 27.90 |
| 1985 Sep 17.5 | 28.5 | 3.68 | −10.26 | −11.57 | −10.52 | −11.24 | −11.30 | −13.47 | ... | −13.17 | 31.21 | 28.35 | 28.68 | 27.58 | 27.87 |
| 1985 Oct 13.5 | 28.5 | 3.79 | −10.42 | −11.85 | −10.79 | −11.36 | −11.67 | −13.77 | ... | −13.33 | 31.15 | 28.35 | 28.67 | 27.76 | 27.80 |
| 1985 Oct 15.5 | 20.2 | 3.65 | −10.63 | −12.30 | −11.05 | −11.74 | −11.82 | −13.80 | ... | −13.48 | 30.93 | 27.94 | 28.45 | 27.41 | 27.68 |
| 1985 Nov  7.5 | 35.3 | 3.98 | −10.29 | −12.53 | −10.84 | −11.66 | −11.57 | −13.66 | ... | −13.40 | 31.40 | 28.05 | 28.98 | 27.80 | 28.24 |
| 1985 Nov 20.5 | 49.8 | 4.18 | −10.16 | −11.94 | −10.75 | −11.45 | −11.54 | −13.76 | ... | −13.88 | 31.68 | 28.80 | 29.23 | 28.16 | 28.42 |
| 1998 Sep  9.2 | 73.7 | 4.50 | −10.39 | −12.23 | −10.37 | −11.17 | −11.30 | −13.68 | und | −13.19 | 32.21 | 28.81 | 30.00 | 28.74 | 28.95 |
| 1998 Sep  9.2 | 51.7 | 4.35 | −10.52 | und | −10.57 | −11.02 | −11.24 | −13.68 | −14.17 | −14.07 | 32.09 | und | 29.80 | 28.89 | 29.01 |
| 1998 Sep 17.1 | 57.6 | 4.38 | −10.27 | −12.03 | −10.41 | −11.25 | −11.05 | −13.88 | −13.55 | −13.55 | 32.22 | 28.95 | 29.89 | 28.59 | 29.14 |
| 1998 Sep 17.2 | 81.4 | 4.53 | −10.04 | −11.79 | −10.19 | −11.13 | −10.84 | −13.77 | −13.44 | −13.46 | 32.44 | 29.19 | 30.10 | 28.71 | 29.35 |
| 1998 Nov  4.1 | 40.4 | 4.12 | −9.98 | −11.46 | −10.21 | −10.90 | −10.84 | −13.25 | −13.07 | −13.02 | 32.20 | 29.08 | 29.65 | 28.52 | 28.93 |
| 1998 Nov  4.1 | 28.7 | 3.98 | −10.18 | −11.63 | −10.46 | −11.34 | −11.00 | −13.29 | −13.11 | −13.33 | 32.00 | 28.91 | 29.40 | 28.08 | 28.76 |
| 1998 Nov 17.1 | 92.7 | 4.46 | −9.47 | −10.87 | −9.94 | −10.80 | −10.34 | −13.09 | −12.84 | −12.82 | 32.68 | 29.70 | 29.97 | 28.56 | 29.36 |
| 1998 Nov 17.1 | 92.7 | 4.46 | −9.47 | −10.87 | −9.93 | −10.91 | −10.33 | −13.04 | −12.82 | −12.81 | 32.68 | 29.71 | 29.98 | 28.44 | 29.37 |
| 1998 Nov 18.1 | 59.4 | 4.27 | −9.77 | −11.16 | −10.21 | −10.93 | −10.62 | −13.33 | −13.00 | −12.99 | 32.39 | 29.41 | 29.71 | 28.42 | 29.08 |
| 1998 Nov 18.1 | 36.7 | 4.06 | −10.10 | −11.52 | −10.51 | −11.14 | −10.93 | −13.54 | −13.18 | −13.16 | 32.06 | 29.06 | 29.41 | 28.21 | 28.77 |
| 1998 Nov 18.2 | 92.7 | 4.46 | −9.47 | −10.86 | −9.95 | −10.83 | −10.35 | −13.18 | −12.86 | −12.84 | 32.68 | 29.71 | 29.96 | 28.52 | 29.34 |
| 1998 Nov 19.1 | 59.4 | 4.27 | −9.79 | −11.14 | −10.21 | −10.92 | −10.62 | −13.28 | −13.01 | −12.97 | 32.38 | 29.43 | 29.72 | 28.42 | 29.07 |
| 1998 Nov 19.1 | 36.7 | 4.06 | −10.11 | −11.49 | −10.49 | −11.13 | −10.92 | −13.52 | −13.18 | −13.13 | 32.06 | 29.08 | 29.44 | 28.22 | 28.78 |
| 1998 Dec  8.1 | 59.4 | 4.27 | −9.97 | −11.56 | −10.32 | −11.24 | −11.02 | −13.66 | −13.38 | −13.36 | 31.99 | 28.88 | 29.39 | 28.14 | 28.71 |
| 1998 Dec  8.1 | 120.7 | 4.58 | −9.48 | −11.00 | −9.89 | −11.00 | −10.58 | −13.47 | −13.10 | −13.09 | 32.48 | 29.44 | 29.81 | 28.37 | 29.14 |
| 1998 Dec 11.1 | 120.7 | 4.58 | −9.57 | −11.09 | −9.95 | −11.30 | −10.65 | −13.36 | −13.12 | −13.10 | 32.40 | 29.36 | 29.77 | 28.10 | 29.09 |
| 1998 Dec 11.1 | 59.4 | 4.28 | −10.04 | −11.65 | −10.36 | −11.46 | −11.08 | −13.65 | −13.40 | −13.37 | 31.93 | 28.80 | 29.36 | 27.94 | 28.67 |
| 1998 Dec 11.1 | 120.7 | 4.58 | −9.56 | −11.09 | −9.95 | −11.19 | −10.64 | −13.31 | −13.11 | −13.11 | 32.40 | 29.36 | 29.76 | 28.20 | 29.10 |

Table 2—Continued

| UT Date | Aperture | | log Emission Band Flux[a] | | | | | log Continuum Flux[a] | | | log $M(\rho)$[a] | | | | |
|---|---|---|---|---|---|---|---|---|---|---|---|---|---|---|---|
| | Size | log $\rho$ | (erg cm$^{-2}$ s$^{-1}$) | | | | | (erg cm$^{-2}$ s$^{-1}$ Å$^{-1}$) | | | (molecule) | | | | |
| | (arcsec) | (km) | OH | NH | CN | C$_3$ | C$_2$ | UV | Blue | Green | OH | NH | CN | C$_3$ | C$_2$ |
| 1998 Dec 20.5 | 146.7 | 4.69 | ... | ... | −9.82 | und | −10.37 | ... | −12.76 | −13.11 | ... | ... | 29.97 | und | 29.44 |
| 1998 Dec 20.5 | 73.7 | 4.39 | ... | ... | −10.14 | −10.78 | −10.79 | ... | −13.25 | −13.25 | ... | ... | 29.64 | 28.69 | 29.02 |
| 1999 Jan 9.1 | 36.7 | 4.15 | −10.21 | −12.24 | −10.75 | und | −11.44 | −13.70 | −13.49 | −13.50 | 31.78 | 28.55 | 29.29 | und | 28.59 |
| 1999 Jan 9.1 | 92.7 | 4.55 | −9.88 | −11.72 | −10.39 | −11.19 | −11.18 | −14.04 | −13.32 | −13.60 | 32.11 | 29.06 | 29.66 | 28.49 | 28.85 |
| 1999 Jan 10.1 | 148.6 | 4.76 | −9.38 | −11.39 | −9.95 | −11.20 | −10.62 | −13.39 | −13.16 | −13.13 | 32.61 | 29.41 | 30.10 | 28.50 | 29.42 |
| 1999 Jan 10.1 | 92.7 | 4.56 | −9.66 | −11.89 | −10.20 | −11.63 | −10.89 | −13.52 | −13.27 | −13.23 | 32.34 | 28.91 | 29.86 | 28.06 | 29.15 |
| 1999 Jan 10.2 | 59.4 | 4.36 | −9.92 | −12.10 | −10.48 | −11.62 | −11.16 | −13.67 | −13.37 | −13.36 | 32.08 | 28.70 | 29.59 | 28.07 | 28.88 |
| 1999 Jan 18.5 | 146.7 | 4.79 | ... | ... | −10.13 | und | −10.86 | ... | −13.07 | −13.11 | ... | ... | 30.04 | und | 29.29 |
| 1999 Jan 18.5 | 73.7 | 4.49 | ... | ... | −10.47 | und | −11.69 | ... | −13.39 | −13.42 | ... | ... | 29.70 | und | 28.46 |
| 1999 Jan 19.5 | 146.7 | 4.79 | ... | ... | −10.12 | und | −10.87 | ... | −13.02 | −13.18 | ... | ... | 30.06 | und | 29.29 |
| 1999 Jan 19.6 | 73.7 | 4.49 | ... | ... | −10.61 | −11.02 | −10.81 | ... | −13.59 | −13.44 | ... | ... | 29.57 | 28.80 | 29.35 |
| 1999 Mar 4.1 | 194.9 | 4.97 | −10.35 | −12.94 | −10.92 | −12.00 | −11.64 | −14.01 | −13.72 | −13.72 | 31.89 | 28.21 | 29.46 | 28.01 | 28.72 |
| 1999 Mar 4.1 | 194.9 | 4.97 | −10.36 | −12.57 | −10.92 | −11.98 | −11.66 | −14.02 | −13.76 | −13.71 | 31.89 | 28.58 | 29.46 | 28.03 | 28.70 |
| 1999 Mar 4.2 | 194.9 | 4.97 | −10.36 | −12.97 | −10.92 | −12.04 | −11.65 | −14.01 | −13.74 | −13.70 | 31.89 | 28.17 | 29.46 | 27.98 | 28.71 |
| 1999 May 8.2 | 148.6 | 5.17 | ... | ... | −12.49 | −12.71 | und | ... | −14.90 | −14.59 | ... | ... | 28.93 | 28.33 | und |
| 1999 May 9.2 | 148.6 | 5.17 | ... | ... | −12.72 | ... | und | ... | ... | −14.46 | ... | ... | 28.71 | ... | und |
| 1999 May 9.2 | 148.6 | 5.17 | ... | ... | −12.39 | ... | und | ... | ... | −14.80 | ... | ... | 29.04 | ... | und |
| 1999 May 9.2 | 148.6 | 5.17 | ... | ... | −12.57 | ... | −12.43 | ... | ... | −15.28 | ... | ... | 28.86 | ... | 28.97 |
| 2011 Oct 20.1 | 97.2 | 4.91 | ... | ... | −11.12 | ... | −11.57 | ... | und | −14.57 | ... | ... | 30.04 | ... | 29.47 |
| 2018 May 17.3 | 97.2 | 4.67 | −11.23 | und | −11.12 | −11.65 | −11.84 | −15.00 | und | −14.48 | 31.78 | und | 29.57 | 28.57 | 28.73 |
| 2018 May 17.3 | 97.2 | 4.67 | −11.36 | und | −11.11 | −12.28 | −11.69 | −14.29 | −14.46 | −15.83 | 31.64 | und | 29.57 | 27.94 | 28.88 |
| 2018 May 17.3 | 77.8 | 4.58 | −11.48 | und | −11.25 | und | −12.38 | und | −14.37 | −14.27 | 31.52 | und | 29.43 | und | 28.19 |
| 2018 May 17.3 | 97.2 | 4.67 | −11.33 | −12.94 | −11.10 | −12.17 | −11.85 | und | −14.64 | −14.56 | 31.68 | 28.41 | 29.58 | 28.05 | 28.72 |
| 2018 May 17.4 | 97.2 | 4.67 | −11.38 | −13.06 | −11.11 | −11.72 | −11.89 | −14.62 | −14.30 | −14.40 | 31.63 | 28.29 | 29.57 | 28.50 | 28.68 |
| 2018 Jun 5.3 | 62.4 | 4.40 | −11.52 | und | −11.48 | −11.85 | −11.75 | und | −14.01 | −14.03 | 31.23 | und | 28.95 | 28.11 | 28.57 |
| 2018 Jun 5.3 | 62.4 | 4.40 | −11.14 | −12.50 | −10.99 | −11.60 | −11.70 | −14.59 | −14.16 | −14.27 | 31.62 | 28.60 | 29.44 | 28.33 | 28.61 |
| 2018 Jun 5.3 | 97.2 | 4.59 | −10.79 | −12.06 | −10.69 | −11.53 | −11.38 | −14.49 | −13.84 | −14.37 | 31.96 | 29.04 | 29.74 | 28.44 | 28.94 |
| 2018 Jun 12.2 | 97.2 | 4.56 | −10.62 | −12.54 | −10.59 | −11.08 | −11.28 | −14.59 | −13.97 | −14.03 | 32.03 | 28.46 | 29.74 | 28.79 | 28.94 |
| 2018 Jun 12.2 | 97.2 | 4.56 | −10.72 | und | −10.59 | −12.11 | −11.24 | −13.81 | −14.23 | −14.03 | 31.93 | und | 29.74 | 27.76 | 28.98 |
| 2018 Jun 12.2 | 77.8 | 4.46 | −10.91 | −12.22 | −10.70 | −11.37 | −11.40 | −14.56 | −13.97 | −13.97 | 31.74 | 28.79 | 29.63 | 28.51 | 28.82 |
| 2018 Jun 12.3 | 97.2 | 4.56 | −10.68 | −12.38 | −10.57 | −11.29 | −11.29 | −14.46 | −13.90 | −13.84 | 31.97 | 28.62 | 29.76 | 28.59 | 28.93 |
| 2018 Jun 12.3 | 77.8 | 4.46 | −10.82 | −12.42 | −10.70 | −11.60 | −11.30 | −14.28 | −13.95 | −14.95 | 31.83 | 28.58 | 29.62 | 28.27 | 28.92 |
| 2018 Jun 12.3 | 62.4 | 4.37 | −10.95 | −12.53 | −10.86 | −11.58 | −11.61 | −14.39 | −14.06 | −14.00 | 31.70 | 28.47 | 29.46 | 28.29 | 28.61 |
| 2018 Jul 3.2 | 48.6 | 4.16 | −10.56 | −12.29 | −10.58 | −11.53 | −11.27 | −13.88 | −13.60 | −13.59 | 31.72 | 28.41 | 29.45 | 28.05 | 28.65 |
| 2018 Jul 3.2 | 62.4 | 4.27 | −10.38 | −12.22 | −10.44 | −11.24 | −11.12 | −13.87 | −13.52 | −13.50 | 31.90 | 28.48 | 29.58 | 28.33 | 28.80 |
| 2018 Jul 3.2 | 97.2 | 4.47 | −10.07 | −11.89 | −10.15 | −11.28 | −10.82 | −13.64 | −13.42 | −13.41 | 32.20 | 28.81 | 29.87 | 28.29 | 29.10 |
| 2018 Jul 5.2 | 62.4 | 4.26 | −10.28 | −12.24 | −10.39 | −11.34 | −11.09 | −13.65 | −13.39 | −13.39 | 31.95 | 28.43 | 29.61 | 28.20 | 28.80 |
| 2018 Jul 5.2 | 77.8 | 4.36 | −10.16 | −11.96 | −10.24 | −11.08 | −10.91 | −13.73 | −13.36 | −13.39 | 32.07 | 28.71 | 29.76 | 28.46 | 28.98 |
| 2018 Jul 5.2 | 62.4 | 4.26 | −10.33 | −12.03 | −10.40 | −11.33 | −11.06 | −13.71 | −13.42 | −13.45 | 31.91 | 28.65 | 29.60 | 28.21 | 28.83 |
| 2018 Aug 5.2 | 97.2 | 4.30 | −9.13 | −10.87 | −9.57 | −10.42 | −10.17 | −12.77 | −12.49 | −12.48 | 32.49 | 29.33 | 29.94 | 28.63 | 29.23 |
| 2018 Aug 5.2 | 155.9 | 4.50 | −8.84 | −10.59 | −9.29 | −10.27 | −9.89 | −12.62 | −12.38 | −12.37 | 32.77 | 29.62 | 30.21 | 28.79 | 29.52 |
| 2018 Aug 5.2 | 62.4 | 4.10 | −9.43 | −11.16 | −9.85 | −10.58 | −10.46 | −12.95 | −12.63 | −12.61 | 32.18 | 29.04 | 29.65 | 28.48 | 28.94 |
| 2018 Aug 5.2 | 48.6 | 4.00 | −9.59 | −11.32 | −10.00 | −10.69 | −10.64 | −13.04 | −12.71 | −12.68 | 32.02 | 28.88 | 29.50 | 28.37 | 28.77 |
| 2018 Aug 5.2 | 77.8 | 4.20 | −9.27 | −11.02 | −9.69 | −10.52 | −10.30 | −12.84 | −12.54 | −12.54 | 32.34 | 29.19 | 29.81 | 28.54 | 29.10 |

Table 2—Continued

| UT Date | Aperture | | log Emission Band Flux[a] | | | | | log Continuum Flux[a] | | | log $M(\rho)$[a] | | | | |
|---|---|---|---|---|---|---|---|---|---|---|---|---|---|---|---|
| | Size | log $\rho$ | (erg cm$^{-2}$ s$^{-1}$) | | | | | (erg cm$^{-2}$ s$^{-1}$ Å$^{-1}$) | | | (molecule) | | | | |
| | (arcsec) | (km) | OH | NH | CN | $C_3$ | $C_2$ | UV | Blue | Green | OH | NH | CN | $C_3$ | $C_2$ |
| 2018 Aug 5.3 | 97.2 | 4.30 | −9.13 | −10.84 | −9.57 | −10.42 | −10.16 | −12.78 | −12.49 | −12.48 | 32.48 | 29.36 | 29.94 | 28.64 | 29.24 |
| 2018 Aug 19.4 | 97.2 | 4.21 | −9.15 | −10.54 | −9.40 | −10.19 | −10.00 | −12.54 | −12.19 | −12.17 | 32.47 | 29.41 | 29.86 | 28.64 | 29.18 |
| 2018 Aug 19.4 | 155.9 | 4.42 | −8.87 | −10.24 | −9.12 | −10.03 | −9.71 | −12.40 | −12.08 | −12.06 | 32.75 | 29.71 | 30.14 | 28.81 | 29.47 |
| 2018 Aug 19.4 | 48.6 | 3.91 | −9.60 | −11.03 | −9.84 | −10.51 | −10.48 | −12.77 | −12.40 | −12.38 | 32.02 | 28.92 | 29.41 | 28.32 | 28.70 |
| 2018 Aug 19.4 | 97.2 | 4.21 | −9.16 | −10.54 | −9.40 | −10.18 | −10.00 | −12.53 | −12.19 | −12.17 | 32.46 | 29.41 | 29.86 | 28.66 | 29.18 |
| 2018 Sep 5.3 | 97.2 | 4.14 | −9.11 | −10.50 | −9.51 | −10.13 | −9.95 | −12.52 | −12.18 | −12.15 | 32.35 | 29.37 | 29.70 | 28.53 | 29.05 |
| 2018 Sep 5.3 | 48.6 | 3.84 | −9.56 | −10.99 | −9.95 | −10.47 | −10.44 | −12.78 | −12.42 | −12.38 | 31.89 | 28.88 | 29.26 | 28.19 | 28.57 |
| 2018 Sep 5.4 | 97.2 | 4.14 | −9.11 | −10.49 | −9.51 | −10.13 | −9.95 | −12.54 | −12.19 | −12.16 | 32.34 | 29.39 | 29.70 | 28.53 | 29.05 |
| 2018 Sep 13.3 | 204.5 | 4.46 | −8.74 | −10.09 | −9.11 | −9.96 | −9.59 | −12.37 | −12.03 | −12.02 | 32.72 | 29.73 | 30.08 | 28.69 | 29.41 |
| 2018 Sep 13.3 | 126.7 | 4.26 | −9.02 | −10.38 | −9.38 | −10.11 | −9.86 | −12.52 | −12.16 | −12.14 | 32.43 | 29.44 | 29.81 | 28.54 | 29.14 |
| 2018 Sep 13.4 | 97.2 | 4.14 | −9.19 | −10.57 | −9.54 | −10.21 | −10.04 | −12.61 | −12.25 | −12.22 | 32.27 | 29.25 | 29.65 | 28.44 | 28.96 |
| 2018 Sep 13.4 | 48.6 | 3.84 | −9.67 | −11.03 | −9.99 | −10.53 | −10.50 | −12.88 | −12.49 | −12.47 | 31.78 | 28.79 | 29.20 | 28.12 | 28.50 |
| 2018 Sep 13.4 | 62.4 | 3.95 | −9.51 | −10.91 | −9.83 | −10.42 | −10.33 | −12.79 | −12.42 | −12.38 | 31.94 | 28.91 | 29.36 | 28.23 | 28.66 |
| 2018 Sep 13.4 | 97.2 | 4.14 | −9.21 | −10.54 | −9.54 | −10.20 | −10.02 | −12.61 | −12.25 | −12.23 | 32.24 | 29.28 | 29.65 | 28.45 | 28.97 |
| 2018 Sep 13.4 | 77.8 | 4.04 | −9.35 | −10.68 | −9.67 | −10.28 | −10.15 | −12.69 | −12.33 | −12.31 | 32.11 | 29.14 | 29.52 | 28.37 | 28.84 |
| 2018 Sep 18.4 | 97.2 | 4.15 | −9.19 | −10.57 | −9.47 | −10.30 | −10.07 | −12.68 | −12.34 | −12.31 | 32.18 | 29.22 | 29.62 | 28.38 | 28.95 |
| 2018 Sep 18.4 | 155.9 | 4.36 | −8.90 | −10.26 | −9.19 | −10.12 | −9.79 | −12.51 | −12.20 | −12.18 | 32.47 | 29.53 | 29.90 | 28.56 | 29.24 |
| 2018 Sep 18.4 | 48.6 | 3.85 | −9.74 | −11.13 | −9.96 | −10.62 | −10.57 | −13.01 | −12.62 | −12.58 | 31.63 | 28.66 | 29.14 | 28.06 | 28.45 |
| 2018 Oct 6.4 | 48.6 | 3.93 | −9.89 | −11.50 | −10.26 | −11.23 | −10.97 | −13.38 | −13.10 | −13.06 | 31.56 | 28.44 | 28.95 | 27.66 | 28.26 |
| 2018 Oct 6.4 | 97.2 | 4.23 | −9.43 | −10.95 | −9.83 | −10.85 | −10.51 | −13.13 | −12.82 | −12.82 | 32.02 | 28.99 | 29.38 | 28.04 | 28.73 |
| 2018 Oct 6.4 | 77.8 | 4.14 | −9.59 | −11.13 | −9.96 | −10.94 | −10.63 | −13.19 | −12.92 | −12.91 | 31.86 | 28.81 | 29.25 | 27.95 | 28.61 |
| 2018 Oct 6.4 | 62.4 | 4.04 | −9.76 | −11.32 | −10.11 | −11.05 | −10.82 | −13.29 | −13.00 | −12.98 | 31.69 | 28.62 | 29.10 | 27.84 | 28.42 |
| 2018 Nov 7.4 | 77.8 | 4.30 | −10.33 | *und* | −10.92 | −11.60 | −11.48 | −13.90 | −13.64 | −13.89 | 31.27 | *und* | 28.82 | 27.77 | 28.24 |
| 2018 Nov 7.4 | 48.6 | 4.09 | −10.60 | −12.73 | −11.20 | −15.48 | −11.95 | −14.12 | −13.93 | −13.85 | 31.01 | 27.77 | 28.54 | 23.89 | 27.76 |
| 2018 Nov 7.5 | 62.4 | 4.20 | −10.39 | −12.24 | −11.05 | −11.82 | −11.75 | −14.82 | −13.86 | −13.90 | 31.22 | 28.25 | 28.69 | 27.55 | 27.96 |
| 2018 Nov 15.4 | 48.6 | 4.13 | −10.73 | *und* | −11.38 | −13.05 | −12.02 | −14.32 | −14.22 | −14.14 | 30.97 | *und* | 28.47 | 26.43 | 27.80 |
| 2018 Nov 15.4 | 48.6 | 4.13 | −10.67 | −13.58 | −11.37 | *und* | −12.01 | −14.40 | −14.04 | −14.11 | 31.04 | 27.03 | 28.47 | *und* | 27.82 |
| 2018 Nov 15.4 | 62.4 | 4.23 | −10.53 | −13.64 | −11.18 | −11.85 | −11.84 | *und* | −14.00 | −14.01 | 31.18 | 26.97 | 28.66 | 27.63 | 27.99 |
| 2018 Nov 15.4 | 77.8 | 4.33 | −10.37 | −12.94 | −11.01 | −11.66 | −11.78 | −14.57 | −13.97 | −13.90 | 31.33 | 27.66 | 28.83 | 27.82 | 28.05 |
| 2018 Dec 16.3 | 97.2 | 4.53 | −11.74 | *und* | −11.32 | −11.30 | −11.97 | *und* | −13.36 | −13.58 | 30.35 | *und* | 28.90 | 28.55 | 28.22 |
| 2018 Dec 16.3 | 77.8 | 4.43 | −10.92 | *und* | −11.61 | −12.93 | −11.93 | | −13.73 | −13.82 | 31.17 | *und* | 28.60 | 26.92 | 28.26 |
| 2018 Dec 16.4 | 77.8 | 4.43 | *und* | −12.41 | −11.48 | −11.95 | −12.45 | −14.55 | −14.03 | −13.98 | *und* | 28.58 | 28.73 | 27.90 | 27.75 |
| 2018 Dec 16.4 | 62.4 | 4.34 | −11.11 | *und* | −11.66 | *und* | −12.15 | −13.98 | −14.14 | −14.27 | 30.98 | *und* | 28.56 | *und* | 28.05 |
| 2018 Dec 30.3 | 97.2 | 4.58 | −10.52 | *und* | −11.51 | *und* | −12.21 | −14.40 | −13.84 | −14.05 | 31.73 | *und* | 28.87 | *und* | 28.14 |
| 2018 Dec 30.3 | 77.8 | 4.58 | −10.61 | *und* | −11.54 | *und* | −11.99 | −14.08 | −14.07 | −14.06 | 31.64 | *und* | 28.84 | *und* | 28.37 |
| 2019 Jan 31.3 | 77.8 | 4.60 | ... | ... | −11.92 | ... | *und* | ... | −14.51 | −14.44 | ... | ... | 28.82 | ... | *und* |

[a] "*und*" stands for "undefined". For the gases, this means that the emission flux was measured but was less than zero following sky and continuum removal. For the continuum, this means the continuum flux was measured but following sky subtraction it was less than zero.

Table 3.  Photometric production rates for comet 21P/Giacobini-Zinner.

| UT Date | ΔT | log $r_H$ | log $ρ$ | log $Q^{a,b}$ (molecules s$^{-1}$) | | | | | log $A(θ)fρ^{a,b}$ (cm) | | | log $Q^a$ |
| | | | | OH | NH | CN | $C_3$ | $C_2$ | UV | Blue | Green | $H_2O$ |
| | (day) | (AU) | (km) | | | | | | | | | |
|---|---|---|---|---|---|---|---|---|---|---|---|---|
| 1985 Jun 15.3 | −83.4 | 0.180 | 3.87 | $28.02_{.02}$ | $25.19_{.10}$ | $25.61_{.00}$ | $24.09_{.05}$ | $24.90_{.03}$ | $2.18_{.03}$ | ... | $2.25_{.01}$ | 28.06 |
| 1985 Jun 15.4 | −83.3 | 0.180 | 3.71 | ... | ... | $25.60_{.01}$ | $24.09_{.07}$ | $24.75_{.06}$ | $2.20_{.03}$ | ... | $2.31_{.01}$ | ... |
| 1985 Jun 16.3 | −82.5 | 0.178 | 3.86 | $28.01_{.02}$ | $25.34_{.09}$ | $25.59_{.01}$ | $24.16_{.05}$ | $24.78_{.05}$ | $2.13_{.03}$ | ... | $2.26_{.01}$ | 28.05 |
| 1985 Jun 16.4 | −82.4 | 0.178 | 3.86 | $28.05_{.01}$ | $25.26_{.09}$ | $25.61_{.00}$ | $24.13_{.05}$ | $24.85_{.03}$ | $2.14_{.03}$ | ... | $2.26_{.01}$ | 28.09 |
| 1985 Jun 16.4 | −82.3 | 0.178 | 3.70 | ... | ... | $25.62_{.01}$ | $24.13_{.06}$ | $24.80_{.05}$ | $2.22_{.03}$ | ... | $2.33_{.01}$ | ... |
| 1985 Aug 11.4 | −26.4 | 0.038 | 3.75 | $28.55_{.00}$ | $25.75_{.01}$ | $25.72_{.00}$ | $24.35_{.01}$ | $25.15_{.01}$ | $2.56_{.00}$ | ... | $2.61_{.00}$ | 28.67 |
| 1985 Aug 11.4 | −26.3 | 0.038 | 3.60 | $28.54_{.00}$ | $25.74_{.01}$ | $25.72_{.00}$ | $24.37_{.01}$ | $25.13_{.01}$ | $2.54_{.01}$ | ... | $2.60_{.00}$ | 28.66 |
| 1985 Aug 12.4 | −25.4 | 0.036 | 3.74 | $28.55_{.00}$ | $25.70_{.01}$ | $25.73_{.00}$ | $24.39_{.01}$ | $25.09_{.01}$ | $2.48_{.01}$ | ... | $2.57_{.00}$ | 28.67 |
| 1985 Aug 12.4 | −25.3 | 0.036 | 3.89 | $28.54_{.00}$ | $25.68_{.01}$ | $25.73_{.00}$ | $24.41_{.01}$ | $25.16_{.00}$ | $2.46_{.00}$ | ... | $2.55_{.00}$ | 28.66 |
| 1985 Aug 12.5 | −25.3 | 0.036 | 3.44 | $28.55_{.01}$ | ... | $25.74_{.00}$ | $24.41_{.02}$ | $25.06_{.02}$ | $2.48_{.01}$ | ... | $2.59_{.00}$ | 28.67 |
| 1985 Aug 13.4 | −24.3 | 0.034 | 3.74 | $28.54_{.00}$ | $25.72_{.01}$ | $25.73_{.00}$ | $24.38_{.01}$ | $25.15_{.01}$ | $2.50_{.00}$ | ... | $2.58_{.00}$ | 28.65 |
| 1985 Aug 20.4 | −17.4 | 0.023 | 3.71 | $28.45_{.00}$ | $25.65_{.01}$ | $25.67_{.00}$ | $24.36_{.01}$ | $25.06_{.01}$ | $2.39_{.01}$ | ... | $2.50_{.00}$ | 28.57 |
| 1985 Sep 9.4 | +2.6 | 0.012 | 3.76 | $28.25_{.01}$ | $25.53_{.04}$ | $25.46_{.00}$ | $24.11_{.04}$ | $24.87_{.02}$ | $2.16_{.02}$ | ... | $2.26_{.01}$ | 28.38 |
| 1985 Sep 9.4 | +2.7 | 0.012 | 3.76 | $28.26_{.01}$ | $25.49_{.03}$ | $25.47_{.00}$ | $24.13_{.05}$ | $24.91_{.01}$ | $2.18_{.02}$ | ... | $2.24_{.01}$ | 28.39 |
| 1985 Sep 9.5 | +2.7 | 0.012 | 3.61 | $28.27_{.01}$ | $25.65_{.03}$ | $25.49_{.00}$ | $24.15_{.03}$ | $24.90_{.01}$ | $2.18_{.02}$ | ... | $2.26_{.01}$ | 28.40 |
| 1985 Sep 9.5 | +2.7 | 0.012 | 3.91 | $28.26_{.00}$ | $25.60_{.02}$ | $25.47_{.00}$ | $24.19_{.02}$ | $24.90_{.01}$ | $2.13_{.02}$ | ... | $2.23_{.01}$ | 28.39 |
| 1985 Sep 12.4 | +5.7 | 0.014 | 3.98 | $28.15_{.01}$ | $25.44_{.04}$ | $25.40_{.00}$ | $23.92_{.05}$ | $24.82_{.01}$ | $2.11_{.03}$ | ... | $2.16_{.01}$ | 28.28 |
| 1985 Sep 12.4 | +5.7 | 0.014 | 3.98 | $28.14_{.01}$ | $25.52_{.03}$ | $25.40_{.00}$ | $24.06_{.04}$ | $24.81_{.01}$ | $2.07_{.03}$ | ... | $2.16_{.01}$ | 28.27 |
| 1985 Sep 12.5 | +5.7 | 0.014 | 3.83 | $28.13_{.01}$ | $25.41_{.04}$ | $25.40_{.00}$ | $24.04_{.04}$ | $24.75_{.02}$ | $2.08_{.03}$ | ... | $2.17_{.01}$ | 28.26 |
| 1985 Sep 13.5 | +6.7 | 0.015 | 3.98 | $28.15_{.01}$ | $25.45_{.03}$ | $25.40_{.00}$ | $23.77_{.06}$ | $24.84_{.01}$ | $2.18_{.02}$ | ... | $2.13_{.01}$ | 28.27 |
| 1985 Sep 16.4 | +9.6 | 0.017 | 3.68 | $28.07_{.01}$ | $25.42_{.02}$ | $25.36_{.00}$ | $24.00_{.02}$ | $24.79_{.01}$ | $2.08_{.01}$ | ... | $2.17_{.00}$ | 28.20 |
| 1985 Sep 17.5 | +10.7 | 0.017 | 3.68 | $28.09_{.01}$ | $25.47_{.01}$ | $25.36_{.00}$ | $23.97_{.02}$ | $24.75_{.01}$ | $2.13_{.01}$ | ... | $2.20_{.00}$ | 28.21 |
| 1985 Oct 13.5 | +36.7 | 0.063 | 3.79 | $27.91_{.01}$ | $25.36_{.03}$ | $25.24_{.00}$ | $24.05_{.03}$ | $24.57_{.03}$ | $2.02_{.02}$ | ... | $2.23_{.00}$ | 28.01 |
| 1985 Oct 15.5 | +38.7 | 0.067 | 3.65 | $27.93_{.01}$ | $25.19_{.05}$ | $25.25_{.00}$ | $23.88_{.04}$ | $24.69_{.02}$ | $2.16_{.01}$ | ... | $2.25_{.00}$ | 28.03 |
| 1985 Nov 7.5 | +61.7 | 0.127 | 3.98 | $27.92_{.02}$ | $24.82_{.30}$ | $25.32_{.01}$ | $23.87_{.14}$ | $24.77_{.03}$ | $2.27_{.05}$ | ... | $2.30_{.01}$ | 27.99 |
| 1985 Nov 20.5 | +74.7 | 0.161 | 4.18 | $27.93_{.03}$ | $25.29_{.14}$ | $25.30_{.01}$ | $24.03_{.12}$ | $24.69_{.06}$ | $2.13_{.07}$ | ... | $2.28_{.01}$ | 27.98 |
| 1998 Sep 9.2 | −73.6 | 0.157 | 4.50 | $27.96_{.03}$ | $24.77_{.15}$ | $25.59_{.00}$ | $24.31_{.05}$ | $24.74_{.04}$ | $2.28_{.04}$ | *und* | $2.46_{.01}$ | 28.01 |
| 1998 Sep 9.2 | −73.6 | 0.157 | 4.35 | $28.07_{.01}$ | *und* | $25.61_{.00}$ | $24.59_{.05}$ | $25.02_{.02}$ | $2.43_{.04}$ | $1.61_{.10}$ | $1.73_{.07}$ | 28.12 |
| 1998 Sep 17.1 | −65.6 | 0.136 | 4.38 | $28.12_{.01}$ | $25.07_{.05}$ | $25.63_{.00}$ | $24.27_{.05}$ | $25.08_{.01}$ | $2.12_{.05}$ | $2.13_{.02}$ | $2.15_{.02}$ | 28.19 |
| 1998 Sep 17.2 | −65.6 | 0.136 | 4.53 | $28.13_{.01}$ | $25.07_{.04}$ | $25.64_{.00}$ | $24.27_{.03}$ | $25.07_{.01}$ | $2.08_{.05}$ | $2.09_{.02}$ | $2.09_{.02}$ | 28.19 |
| 1998 Nov 4.1 | −17.7 | 0.027 | 4.12 | $28.39_{.01}$ | $25.48_{.06}$ | $25.67_{.00}$ | $24.45_{.05}$ | $25.14_{.02}$ | $2.59_{.05}$ | $2.44_{.02}$ | $2.51_{.02}$ | 28.51 |
| 1998 Nov 4.1 | −17.6 | 0.027 | 3.98 | $28.41_{.01}$ | $25.56_{.06}$ | $25.64_{.00}$ | $24.14_{.11}$ | $25.20_{.02}$ | $2.70_{.05}$ | $2.55_{.02}$ | $2.55_{.02}$ | 28.53 |
| 1998 Nov 17.1 | −4.7 | 0.015 | 4.46 | $28.36_{.01}$ | $25.57_{.01}$ | $25.53_{.00}$ | $24.25_{.03}$ | $25.10_{.01}$ | $2.34_{.03}$ | $2.27_{.01}$ | $2.31_{.01}$ | 28.49 |
| 1998 Nov 17.1 | −4.6 | 0.015 | 4.46 | $28.37_{.01}$ | $25.58_{.01}$ | $25.54_{.00}$ | $24.14_{.04}$ | $25.12_{.01}$ | $2.40_{.02}$ | $2.29_{.01}$ | $2.32_{.01}$ | 28.50 |
| 1998 Nov 18.1 | −3.7 | 0.015 | 4.27 | $28.34_{.00}$ | $25.57_{.01}$ | $25.51_{.00}$ | $24.24_{.01}$ | $25.07_{.00}$ | $2.30_{.01}$ | $2.29_{.01}$ | $2.33_{.01}$ | 28.47 |
| 1998 Nov 18.1 | −3.7 | 0.015 | 4.06 | $28.33_{.00}$ | $25.55_{.01}$ | $25.51_{.00}$ | $24.20_{.02}$ | $25.06_{.01}$ | $2.29_{.02}$ | $2.33_{.01}$ | $2.37_{.01}$ | 28.45 |
| 1998 Nov 18.2 | −3.6 | 0.015 | 4.46 | $28.37_{.01}$ | $25.58_{.02}$ | $25.52_{.00}$ | $24.22_{.04}$ | $25.09_{.01}$ | $2.25_{.04}$ | $2.25_{.01}$ | $2.28_{.01}$ | 28.50 |
| 1998 Nov 19.1 | −2.7 | 0.015 | 4.27 | $28.33_{.00}$ | $25.59_{.01}$ | $25.53_{.00}$ | $24.24_{.02}$ | $25.07_{.00}$ | $2.34_{.01}$ | $2.29_{.01}$ | $2.34_{.01}$ | 28.46 |
| 1998 Nov 19.1 | −2.7 | 0.015 | 4.06 | $28.32_{.00}$ | $25.57_{.01}$ | $25.54_{.00}$ | $24.20_{.02}$ | $25.07_{.01}$ | $2.31_{.02}$ | $2.32_{.01}$ | $2.39_{.01}$ | 28.45 |
| 1998 Dec 8.1 | +16.3 | 0.025 | 4.27 | $27.95_{.01}$ | $25.05_{.04}$ | $25.20_{.00}$ | $23.94_{.05}$ | $24.71_{.01}$ | $1.99_{.06}$ | $1.94_{.02}$ | $1.99_{.02}$ | 28.07 |
| 1998 Dec 8.1 | +16.4 | 0.025 | 4.58 | $28.02_{.02}$ | $25.15_{.04}$ | $25.24_{.00}$ | $24.00_{.06}$ | $24.76_{.01}$ | $1.87_{.10}$ | $1.91_{.02}$ | $1.95_{.02}$ | 28.14 |
| 1998 Dec 11.1 | +19.3 | 0.029 | 4.58 | $27.93_{.00}$ | $25.06_{.02}$ | $25.19_{.00}$ | $23.72_{.07}$ | $24.70_{.01}$ | $2.00_{.04}$ | $1.91_{.01}$ | $1.94_{.01}$ | 28.05 |
| 1998 Dec 11.1 | +19.3 | 0.029 | 4.28 | $27.88_{.00}$ | $24.97_{.03}$ | $25.16_{.00}$ | $23.74_{.04}$ | $24.66_{.01}$ | $2.01_{.04}$ | $1.93_{.01}$ | $1.98_{.01}$ | 28.00 |
| 1998 Dec 11.1 | +19.4 | 0.029 | 4.58 | $27.94_{.00}$ | $25.06_{.03}$ | $25.19_{.00}$ | $23.83_{.04}$ | $24.71_{.01}$ | $2.05_{.05}$ | $1.92_{.02}$ | $1.93_{.01}$ | 28.06 |
| 1998 Dec 20.5 | +28.8 | 0.046 | 4.69 | ... | ... | $25.28_{.00}$ | *und* | $24.94_{.01}$ | ... | $2.24_{.01}$ | $1.91_{.01}$ | ... |

Table 3—Continued

| UT Date | $\Delta T$ | log $r_H$ | log $\rho$ | log $Q^{a,b}$ (molecules s$^{-1}$) | | | | | log $A(\theta)f\rho^{a,b}$ (cm) | | | log $Q^a$ |
| --- | --- | --- | --- | --- | --- | --- | --- | --- | --- | --- | --- | --- |
| | (day) | (AU) | (km) | OH | NH | CN | C$_3$ | C$_2$ | UV | Blue | Green | H$_2$O |
| 1998 Dec 20.5 | +28.8 | 0.046 | 4.39 | ... | ... | $25.31_{.00}$ | $24.40_{.02}$ | $24.88_{.01}$ | ... | $2.04_{.01}$ | $2.06_{.01}$ | ... |
| 1999 Jan 9.1 | +48.4 | 0.091 | 4.15 | $27.99_{.01}$ | $24.99_{.06}$ | $25.33_{.00}$ | *und* | $24.83_{.02}$ | $2.37_{.03}$ | $2.25_{.01}$ | $2.27_{.01}$ | 28.08 |
| 1999 Jan 9.1 | +48.4 | 0.091 | 4.55 | $27.73_{.01}$ | $24.88_{.04}$ | $25.14_{.00}$ | $24.07_{.05}$ | $24.53_{.01}$ | $1.64_{.09}$ | $2.03_{.01}$ | $1.77_{.02}$ | 27.82 |
| 1999 Jan 10.1 | +49.3 | 0.094 | 4.76 | $27.96_{.00}$ | $24.93_{.03}$ | $25.35_{.00}$ | $23.97_{.06}$ | $24.85_{.01}$ | $2.09_{.04}$ | $1.99_{.02}$ | $2.04_{.01}$ | 28.05 |
| 1999 Jan 10.1 | +49.4 | 0.094 | 4.56 | $27.96_{.01}$ | $24.72_{.06}$ | $25.34_{.00}$ | $23.64_{.10}$ | $24.82_{.01}$ | $2.17_{.05}$ | $2.08_{.02}$ | $2.12_{.01}$ | 28.04 |
| 1999 Jan 10.2 | +49.4 | 0.094 | 4.36 | $27.97_{.01}$ | $24.81_{.08}$ | $25.33_{.00}$ | $23.78_{.08}$ | $24.81_{.01}$ | $2.21_{.04}$ | $2.18_{.01}$ | $2.21_{.01}$ | 28.05 |
| 1999 Jan 18.5 | +57.8 | 0.116 | 4.79 | ... | ... | $25.26_{.00}$ | *und* | $24.70_{.01}$ | ... | $2.16_{.01}$ | $2.14_{.01}$ | ... |
| 1999 Jan 18.5 | +57.8 | 0.116 | 4.49 | ... | ... | $25.28_{.00}$ | *und* | $24.23_{.05}$ | ... | $2.14_{.02}$ | $2.13_{.02}$ | ... |
| 1999 Jan 19.5 | +58.8 | 0.119 | 4.79 | ... | ... | $25.27_{.00}$ | *und* | $24.70_{.01}$ | ... | $2.22_{.01}$ | $2.08_{.01}$ | ... |
| 1999 Jan 19.6 | +58.8 | 0.119 | 4.49 | ... | ... | $25.15_{.00}$ | $24.40_{.03}$ | $25.12_{.01}$ | ... | $1.95_{.03}$ | $2.13_{.02}$ | ... |
| 1999 Mar 4.1 | +74.3 | 0.160 | 4.97 | $27.01_{.00}$ | $23.48_{.13}$ | $24.48_{.00}$ | $23.33_{.03}$ | $23.95_{.01}$ | $1.58_{.03}$ | $1.54_{.01}$ | $1.56_{.01}$ | 27.07 |
| 1999 Mar 4.1 | +74.4 | 0.160 | 4.97 | $27.01_{.00}$ | $23.85_{.07}$ | $24.48_{.00}$ | $23.35_{.06}$ | $23.92_{.01}$ | $1.56_{.03}$ | $1.50_{.01}$ | $1.57_{.01}$ | 27.06 |
| 1999 Mar 4.2 | +74.4 | 0.160 | 4.97 | $27.01_{.01}$ | $23.44_{.16}$ | $24.48_{.00}$ | $23.29_{.08}$ | $23.93_{.01}$ | $1.57_{.04}$ | $1.52_{.01}$ | $1.58_{.01}$ | 27.06 |
| 1999 May 8.2 | +167.4 | 0.358 | 5.17 | ... | ... | $23.78_{.24}$ | $23.33_{.58}$ | *und* | ... | $1.19_{.42}$ | $1.52_{.19}$ | ... |
| 1999 May 9.2 | +168.4 | 0.360 | 5.17 | ... | ... | $23.56_{.32}$ | ... | *und* | ... | $1.63_{.21}$ | ... | ... |
| 1999 May 9.2 | +168.4 | 0.360 | 5.17 | ... | ... | $23.89_{.18}$ | ... | *und* | ... | $1.30_{.37}$ | ... | ... |
| 1999 May 9.2 | +168.4 | 0.360 | 5.17 | ... | ... | $23.71_{.26}$ | ... | $24.00_{.13}$ | ... | $0.81_{.64}$ | ... | ... |
| 2011 Oct 20.1 | −114.6 | 0.257 | 4.91 | ... | ... | $25.15_{.02}$ | ... | $24.78_{.07}$ | ... | $1.45_{.36}$ | ... | ... |
| 2018 May 17.3 | −116.0 | 0.260 | 4.67 | $27.37_{.11}$ | *und* | $25.00_{.02}$ | $23.96_{.13}$ | $24.35_{.07}$ | $1.09_{.66}$ | *und* | $1.31_{.17}$ | 27.37 |
| 2018 May 17.3 | −116.0 | 0.260 | 4.67 | $27.23_{.09}$ | *und* | $25.00_{.03}$ | $23.33_{.34}$ | $24.50_{.10}$ | $1.80_{.19}$ | $1.31_{.20}$ | $-0.05_{.99}$ | 27.23 |
| 2018 May 17.3 | −116.0 | 0.260 | 4.58 | $27.26_{.08}$ | *und* | $25.00_{.01}$ | *und* | $23.95_{.19}$ | *und* | $1.49_{.13}$ | $1.61_{.10}$ | 27.26 |
| 2018 May 17.3 | −115.9 | 0.260 | 4.67 | $27.27_{.05}$ | $24.21_{.33}$ | $25.01_{.00}$ | $23.44_{.23}$ | $24.35_{.08}$ | *und* | $1.13_{.23}$ | $1.20_{.21}$ | 27.27 |
| 2018 May 17.4 | −115.9 | 0.260 | 4.67 | $27.21_{.05}$ | $24.10_{.40}$ | $25.00_{.03}$ | $23.89_{.12}$ | $24.30_{.08}$ | $1.48_{.26}$ | $1.47_{.12}$ | $1.39_{.14}$ | 27.22 |
| 2018 Jun 5.3 | −97.0 | 0.215 | 4.40 | $27.19_{.12}$ | *und* | $24.73_{.03}$ | $23.76_{.23}$ | $24.55_{.06}$ | *und* | $1.78_{.07}$ | $1.78_{.06}$ | 27.21 |
| 2018 Jun 5.3 | −97.0 | 0.215 | 4.40 | $27.58_{.03}$ | $24.79_{.17}$ | $25.22_{.01}$ | $24.01_{.09}$ | $24.60_{.05}$ | $1.52_{.25}$ | $1.63_{.08}$ | $1.54_{.10}$ | 27.60 |
| 2018 Jun 5.3 | −97.0 | 0.215 | 4.59 | $27.63_{.02}$ | $24.92_{.10}$ | $25.25_{.01}$ | $23.91_{.14}$ | $24.64_{.04}$ | $1.43_{.28}$ | $1.76_{.06}$ | $1.25_{.16}$ | 27.65 |
| 2018 Jun 12.2 | −90.1 | 0.198 | 4.56 | $27.72_{.07}$ | $24.37_{.37}$ | $25.28_{.02}$ | $24.30_{.04}$ | $24.67_{.03}$ | $1.27_{.47}$ | $1.56_{.09}$ | $1.52_{.08}$ | 27.76 |
| 2018 Jun 12.2 | −90.1 | 0.198 | 4.56 | $27.63_{.06}$ | *und* | $25.28_{.01}$ | $23.27_{.34}$ | $24.71_{.03}$ | $2.05_{.11}$ | $1.30_{.17}$ | $1.52_{.09}$ | 27.66 |
| 2018 Jun 12.2 | −90.0 | 0.198 | 4.46 | $27.58_{.04}$ | $24.85_{.13}$ | $25.31_{.01}$ | $24.09_{.08}$ | $24.69_{.03}$ | $1.39_{.28}$ | $1.65_{.06}$ | $1.67_{.06}$ | 27.61 |
| 2018 Jun 12.3 | −90.0 | 0.198 | 4.56 | $27.66_{.02}$ | $24.53_{.19}$ | $25.30_{.01}$ | $24.09_{.08}$ | $24.66_{.03}$ | $1.39_{.27}$ | $1.63_{.07}$ | $1.71_{.05}$ | 27.70 |
| 2018 Jun 12.3 | −90.0 | 0.198 | 4.46 | $27.67_{.02}$ | $24.65_{.15}$ | $25.30_{.01}$ | $23.86_{.09}$ | $24.79_{.03}$ | $1.67_{.14}$ | $1.67_{.06}$ | $0.70_{.40}$ | 27.70 |
| 2018 Jun 12.3 | −90.0 | 0.198 | 4.37 | $27.69_{.02}$ | $24.69_{.14}$ | $25.28_{.01}$ | $23.97_{.02}$ | $24.62_{.04}$ | $1.66_{.14}$ | $1.66_{.06}$ | $1.74_{.05}$ | 27.73 |
| 2018 Jul 3.2 | −69.1 | 0.143 | 4.16 | $27.97_{.02}$ | $24.89_{.09}$ | $25.52_{.00}$ | $23.93_{.06}$ | $24.92_{.02}$ | $2.07_{.05}$ | $2.02_{.02}$ | $2.06_{.02}$ | 28.03 |
| 2018 Jul 3.2 | −69.1 | 0.143 | 4.27 | $27.98_{.01}$ | $24.79_{.06}$ | $25.49_{.00}$ | $24.10_{.04}$ | $24.90_{.02}$ | $1.98_{.05}$ | $1.99_{.02}$ | $2.03_{.02}$ | 28.04 |
| 2018 Jul 3.2 | −69.0 | 0.143 | 4.47 | $27.99_{.01}$ | $24.81_{.06}$ | $25.50_{.00}$ | $23.89_{.05}$ | $24.92_{.01}$ | $2.01_{.05}$ | $1.91_{.03}$ | $1.93_{.02}$ | 28.05 |
| 2018 Jul 5.2 | −67.1 | 0.137 | 4.26 | $28.03_{.00}$ | $24.75_{.06}$ | $25.52_{.00}$ | $23.98_{.05}$ | $24.91_{.01}$ | $2.18_{.04}$ | $2.10_{.02}$ | $2.13_{.02}$ | 28.10 |
| 2018 Jul 5.2 | −67.1 | 0.137 | 4.36 | $28.00_{.01}$ | $24.87_{.07}$ | $25.53_{.00}$ | $24.16_{.04}$ | $24.95_{.01}$ | $2.08_{.04}$ | $2.04_{.02}$ | $2.04_{.02}$ | 28.07 |
| 2018 Jul 5.2 | −67.1 | 0.137 | 4.26 | $27.99_{.01}$ | $24.96_{.06}$ | $25.51_{.00}$ | $23.99_{.05}$ | $24.94_{.01}$ | $2.11_{.04}$ | $2.07_{.02}$ | $2.06_{.02}$ | 28.06 |
| 2018 Aug 5.2 | −36.1 | 0.055 | 4.30 | $28.44_{.00}$ | $25.49_{.01}$ | $25.73_{.00}$ | $24.41_{.01}$ | $25.22_{.00}$ | $2.53_{.01}$ | $2.49_{.00}$ | $2.52_{.00}$ | 28.54 |
| 2018 Aug 5.2 | −36.1 | 0.055 | 4.50 | $28.43_{.00}$ | $25.47_{.01}$ | $25.74_{.00}$ | $24.43_{.01}$ | $25.24_{.00}$ | $2.48_{.01}$ | $2.39_{.00}$ | $2.42_{.00}$ | 28.53 |
| 2018 Aug 5.2 | −36.1 | 0.055 | 4.10 | $28.43_{.00}$ | $25.51_{.01}$ | $25.73_{.00}$ | $24.42_{.01}$ | $25.21_{.00}$ | $2.55_{.01}$ | $2.54_{.00}$ | $2.58_{.00}$ | 28.53 |
| 2018 Aug 5.2 | −36.0 | 0.055 | 4.00 | $28.44_{.00}$ | $25.54_{.01}$ | $25.74_{.00}$ | $24.42_{.01}$ | $25.20_{.00}$ | $2.57_{.01}$ | $2.57_{.00}$ | $2.62_{.00}$ | 28.54 |
| 2018 Aug 5.2 | −36.0 | 0.055 | 4.20 | $28.44_{.00}$ | $25.50_{.01}$ | $25.75_{.00}$ | $24.39_{.01}$ | $25.23_{.00}$ | $2.56_{.01}$ | $2.53_{.00}$ | $2.55_{.00}$ | 28.54 |
| 2018 Aug 5.3 | −36.0 | 0.055 | 4.30 | $28.43_{.00}$ | $25.52_{.01}$ | $25.74_{.00}$ | $24.41_{.01}$ | $25.23_{.00}$ | $2.53_{.01}$ | $2.49_{.00}$ | $2.52_{.00}$ | 28.53 |
| 2018 Aug 19.4 | −21.9 | 0.025 | 4.21 | $28.51_{.00}$ | $25.67_{.00}$ | $25.74_{.00}$ | $24.49_{.01}$ | $25.26_{.00}$ | $2.62_{.00}$ | $2.64_{.00}$ | $2.69_{.00}$ | 28.63 |

## Table 3—Continued

| UT Date | ΔT (day) | log $r_H$ (AU) | log $\rho$ (km) | OH | NH | CN | $C_3$ | $C_2$ | UV | Blue | Green | $H_2O$ |
|---|---|---|---|---|---|---|---|---|---|---|---|---|
| | | | | \multicolumn log $Q^{a,b}$ (molecules s$^{-1}$) | | | | | log $A(\theta)f\rho^{a,b}$ (cm) | | | log $Q^a$ |
| 2018 Aug 19.4 | −21.9 | 0.025 | 4.42 | $28.50_{.00}$ | $25.66_{.00}$ | $25.76_{.00}$ | $24.52_{.01}$ | $25.27_{.00}$ | $2.56_{.00}$ | $2.55_{.00}$ | $2.60_{.00}$ | 28.62 |
| 2018 Aug 19.4 | −21.9 | 0.025 | 3.91 | $28.53_{.00}$ | $25.67_{.01}$ | $25.74_{.00}$ | $24.45_{.01}$ | $25.23_{.01}$ | $2.70_{.00}$ | $2.73_{.00}$ | $2.78_{.00}$ | 28.65 |
| 2018 Aug 19.4 | −21.9 | 0.025 | 4.21 | $28.50_{.00}$ | $25.67_{.00}$ | $25.74_{.00}$ | $24.51_{.01}$ | $25.26_{.00}$ | $2.63_{.00}$ | $2.65_{.00}$ | $2.69_{.00}$ | 28.63 |
| 2018 Sep 5.3 | −5.0 | 0.006 | 4.14 | $28.47_{.00}$ | $25.72_{.01}$ | $25.67_{.00}$ | $24.44_{.01}$ | $25.21_{.00}$ | $2.54_{.01}$ | $2.55_{.00}$ | $2.60_{.00}$ | 28.60 |
| 2018 Sep 5.3 | −4.9 | 0.006 | 3.84 | $28.49_{.01}$ | $25.72_{.01}$ | $25.67_{.00}$ | $24.39_{.01}$ | $25.18_{.00}$ | $2.58_{.01}$ | $2.61_{.00}$ | $2.67_{.00}$ | 28.62 |
| 2018 Sep 5.4 | −4.9 | 0.006 | 4.14 | $28.47_{.00}$ | $25.73_{.01}$ | $25.67_{.00}$ | $24.44_{.01}$ | $25.22_{.00}$ | $2.51_{.01}$ | $2.54_{.00}$ | $2.59_{.00}$ | 28.60 |
| 2018 Sep 13.3 | +3.1 | 0.006 | 4.46 | $28.39_{.01}$ | $25.58_{.01}$ | $25.63_{.00}$ | $24.40_{.01}$ | $25.16_{.00}$ | $2.36_{.01}$ | $2.37_{.00}$ | $2.41_{.00}$ | 28.52 |
| 2018 Sep 13.3 | +3.1 | 0.006 | 4.26 | $28.40_{.01}$ | $25.61_{.01}$ | $25.63_{.00}$ | $24.37_{.01}$ | $25.15_{.00}$ | $2.42_{.01}$ | $2.45_{.00}$ | $2.49_{.00}$ | 28.53 |
| 2018 Sep 13.4 | +3.1 | 0.006 | 4.14 | $28.40_{.00}$ | $25.61_{.01}$ | $25.62_{.00}$ | $24.36_{.01}$ | $25.13_{.00}$ | $2.45_{.01}$ | $2.47_{.00}$ | $2.52_{.00}$ | 28.53 |
| 2018 Sep 13.4 | +3.1 | 0.006 | 3.84 | $28.38_{.01}$ | $25.64_{.01}$ | $25.62_{.00}$ | $24.32_{.01}$ | $25.12_{.00}$ | $2.47_{.01}$ | $2.54_{.00}$ | $2.58_{.00}$ | 28.52 |
| 2018 Sep 13.4 | +3.1 | 0.006 | 3.95 | $28.37_{.00}$ | $25.58_{.01}$ | $25.61_{.00}$ | $24.32_{.01}$ | $25.11_{.00}$ | $2.46_{.01}$ | $2.50_{.00}$ | $2.56_{.00}$ | 28.50 |
| 2018 Sep 13.4 | +3.1 | 0.006 | 4.14 | $28.38_{.00}$ | $25.63_{.01}$ | $25.62_{.00}$ | $24.37_{.01}$ | $25.14_{.00}$ | $2.44_{.01}$ | $2.47_{.00}$ | $2.52_{.00}$ | 28.51 |
| 2018 Sep 13.4 | +3.1 | 0.006 | 4.04 | $28.39_{.00}$ | $25.64_{.01}$ | $25.63_{.00}$ | $24.37_{.01}$ | $25.15_{.00}$ | $2.46_{.01}$ | $2.50_{.00}$ | $2.53_{.00}$ | 28.52 |
| 2018 Sep 18.4 | +8.1 | 0.009 | 4.15 | $28.30_{.00}$ | $25.56_{.01}$ | $25.59_{.00}$ | $24.28_{.01}$ | $25.10_{.00}$ | $2.39_{.01}$ | $2.40_{.00}$ | $2.45_{.00}$ | 28.43 |
| 2018 Sep 18.4 | +8.1 | 0.009 | 4.36 | $28.30_{.00}$ | $25.55_{.00}$ | $25.59_{.00}$ | $24.32_{.01}$ | $25.11_{.00}$ | $2.35_{.01}$ | $2.33_{.00}$ | $2.38_{.00}$ | 28.43 |
| 2018 Sep 18.4 | +8.1 | 0.009 | 3.85 | $28.22_{.00}$ | $25.49_{.01}$ | $25.54_{.00}$ | $24.26_{.01}$ | $25.06_{.01}$ | $2.36_{.01}$ | $2.42_{.01}$ | $2.48_{.00}$ | 28.35 |
| 2018 Oct 6.4 | +26.1 | 0.033 | 3.93 | $28.05_{.01}$ | $25.16_{.03}$ | $25.25_{.00}$ | $23.77_{.05}$ | $24.77_{.01}$ | $2.12_{.05}$ | $2.08_{.01}$ | $2.13_{.01}$ | 28.16 |
| 2018 Oct 6.4 | +26.1 | 0.033 | 4.23 | $28.04_{.01}$ | $25.22_{.02}$ | $25.25_{.00}$ | $23.87_{.04}$ | $24.79_{.01}$ | $2.07_{.03}$ | $2.06_{.01}$ | $2.08_{.01}$ | 28.16 |
| 2018 Oct 6.4 | +26.1 | 0.033 | 4.14 | $28.02_{.01}$ | $25.20_{.02}$ | $25.25_{.00}$ | $23.86_{.03}$ | $24.80_{.01}$ | $2.10_{.02}$ | $2.05_{.01}$ | $2.08_{.01}$ | 28.14 |
| 2018 Oct 6.4 | +26.1 | 0.033 | 4.04 | $28.01_{.01}$ | $25.17_{.02}$ | $25.24_{.00}$ | $23.84_{.03}$ | $24.76_{.01}$ | $2.10_{.02}$ | $2.07_{.01}$ | $2.10_{.01}$ | 28.12 |
| 2018 Nov 7.4 | +58.1 | 0.113 | 4.30 | $27.28_{.05}$ | und | $24.66_{.01}$ | $23.52_{.16}$ | $24.28_{.05}$ | $1.71_{.15}$ | $1.65_{.04}$ | $1.42_{.06}$ | 27.36 |
| 2018 Nov 7.4 | +58.2 | 0.113 | 4.09 | $27.34_{.04}$ | $24.33_{.02}$ | $24.68_{.01}$ | $19.84_{.99}$ | $24.11_{.07}$ | $1.70_{.12}$ | $1.56_{.05}$ | $1.66_{.03}$ | 27.41 |
| 2018 Nov 7.5 | +58.2 | 0.113 | 4.20 | $27.37_{.02}$ | $24.64_{.17}$ | $24.67_{.02}$ | $23.39_{.31}$ | $24.16_{.03}$ | $0.89_{.47}$ | $1.52_{.05}$ | $1.51_{.05}$ | 27.45 |
| 2018 Nov 15.4 | +66.1 | 0.135 | 4.13 | $27.28_{.08}$ | und | $24.58_{.02}$ | $22.35_{.80}$ | $24.12_{.08}$ | $1.58_{.19}$ | $1.35_{.09}$ | $1.45_{.07}$ | 27.34 |
| 2018 Nov 15.4 | +66.1 | 0.135 | 4.13 | $27.34_{.06}$ | $23.57_{.82}$ | $24.59_{.02}$ | und | $24.14_{.08}$ | $1.50_{.22}$ | $1.53_{.06}$ | $1.48_{.06}$ | 27.41 |
| 2018 Nov 15.4 | +66.2 | 0.135 | 4.23 | $27.31_{.05}$ | $23.33_{.99}$ | $24.62_{.02}$ | $23.44_{.23}$ | $24.14_{.07}$ | und | $1.46_{.07}$ | $1.48_{.06}$ | 27.37 |
| 2018 Nov 15.4 | +66.2 | 0.135 | 4.33 | $27.31_{.04}$ | $23.87_{.50}$ | $24.64_{.02}$ | $23.54_{.19}$ | $24.06_{.07}$ | $1.12_{.38}$ | $1.39_{.08}$ | $1.49_{.06}$ | 27.38 |
| 2018 Dec 16.3 | +97.0 | 0.216 | 4.53 | $26.11_{.94}$ | und | $24.49_{.09}$ | $24.07_{.27}$ | $24.01_{.26}$ | und | $2.18_{.06}$ | $1.98_{.06}$ | 26.13 |
| 2018 Dec 16.3 | +97.1 | 0.216 | 4.43 | $27.07_{.32}$ | und | $24.34_{.06}$ | $22.53_{.99}$ | $24.20_{.13}$ | und | $1.90_{.06}$ | $1.83_{.07}$ | 27.10 |
| 2018 Dec 16.4 | +97.1 | 0.216 | 4.43 | und | $24.71_{.31}$ | $24.47_{.04}$ | $23.51_{.35}$ | $23.68_{.27}$ | $1.41_{.43}$ | $1.60_{.10}$ | $1.67_{.07}$ | und |
| 2018 Dec 16.4 | +97.1 | 0.216 | 4.34 | $27.04_{.52}$ | und | $24.44_{.03}$ | und | $24.12_{.15}$ | $2.07_{.16}$ | $1.58_{.11}$ | $1.48_{.12}$ | 27.07 |
| 2018 Dec 30.3 | +111.0 | 0.248 | 4.58 | $27.46_{.13}$ | und | $24.43_{.04}$ | und | $23.90_{.21}$ | $1.57_{.35}$ | $1.81_{.08}$ | $1.61_{.10}$ | 27.47 |
| 2018 Dec 30.3 | +111.0 | 0.248 | 4.58 | $27.37_{.14}$ | und | $24.40_{.03}$ | und | $24.12_{.13}$ | $1.89_{.19}$ | $1.57_{.12}$ | $1.61_{.10}$ | 27.38 |
| 2019 Jan 31.3 | +143.0 | 0.315 | 4.60 | ... | ... | $24.42_{.02}$ | ... | und | ... | $1.48_{.13}$ | $1.57_{.11}$ | ... |

[a] *"und"* stands for "undefined". For the gases, this means that the emission flux was measured but was less than zero following sky and continuum removal. For the continuum, this means the continuum flux was measured but following sky subtraction it was less than zero.

[b] Production rates followed by the upper, i.e. the positive, uncertainty. The "+" and "−" uncertainties are equal as percentages, but unequal in log-space; the "−" values can be computed.

Table 4.   Abundance Ratios for Comet 21P/Giacobini-Zinner and Representative Compositional Classes.

| Comet or Class | log Production Rate Ratio | | | | | | | |
|---|---|---|---|---|---|---|---|---|
| | NH/OH | CN/OH | $C_3$/OH | $C_2$/OH | NH/CN | $C_3$/CN | $C_2$/CN | $C_3$/$C_2$ |
| G-Z[a] | $-2.91$ .02 | $-2.58$ .04 | $-3.71$ .15 | $-3.12$ .04 | $-0.20$ .03 | $-1.29$ .04 | $-0.52$ .02 | $-0.77$ .03 |
| Range[b] | | | | | | | | |
| C-C Strong | $-3.29\backslash-2.18$ | $-2.75\backslash-2.24$ | $-4.33\backslash-3.46$ | $-3.86\backslash-3.12$ | $-0.72\backslash+0.27$ | $-1.68\backslash-0.96$ | $-1.12\backslash-0.52$ | $-0.89\backslash+0.32$ |
| C-C Moderate | $-2.54\backslash-1.92$ | $-2.83\backslash-2.41$ | $-3.88\backslash-3.24$ | $-3.35\backslash-2.61$ | $-0.03\backslash+0.74$ | $-1.08\backslash-0.83$ | $-0.72\backslash-0.16$ | $-0.90\backslash-0.49$ |
| Typical | $-2.79\backslash-1.72$ | $-3.00\backslash-2.10$ | $-3.44\backslash-2.58$ | $-2.84\backslash-2.00$ | $-0.52\backslash+0.68$ | $-0.81\backslash-0.24$ | $-0.13\backslash+0.31$ | $-0.83\backslash+0.15$ |
| mean | $-2.25$ | $-2.53$ | $-3.06$ | $-2.43$ | $+0.32$ | $-0.52$ | $+0.11$ | $-0.60$ |
| G-Z Depletion[c] | 4.6× | 1.12× | 4.5× | 4.9× | 3.3× | 5.9× | 4.4× | 1.5× |

[a]The log production rate ratios followed by the upper, i.e. the positive, uncertainty. The "+" and "−" uncertainties are equal as percentages, but unequal in log-space; the "−" values can be computed.

[b]Minimum and maximum values in log-space for members of the strongly carbon-chain depleted, moderately carbon-chain depleted, and typical comet compositional classes as defined by Schleicher & Bair (2016) and amended by Bair & Schleicher (2021), and mean values for the typical compositional class.

[c]Depletion factor of G-Z as compared to the mean typical value.